\documentclass[a4paper,10pt,showkeys]{article}
\usepackage{enumerate,amsmath,amscd,amssymb,graphicx
}

\newcommand{\be}{\begin{equation}}
\newcommand{\ee}{\end{equation}}
\newcommand{\bea}{\begin{eqnarray}}
\newcommand{\eea}{\end{eqnarray}}

\makeatletter
\@addtoreset{equation}{section}

\renewcommand{\thefootnote}{\fnsymbol{footnote}}
\makeatother

\def\del{\delta}

\def\del{\delta}

\def\vep{\varepsilon}

\def\sig{\sigma}

\newcommand{\bsig}{\bar{\sigma}}
\newcommand{\half}{\frac{1}{2}}
\newcommand{\bchi}{\bar{\chi}}

\newcommand{\bpsi}{\bar{\psi}}

\def\del{\partial}
\def\del5{\partial_5}

\begin{document}

\thispagestyle{empty}
%\begin{titlepage}
%
\begin{flushright}
TIT/HEP--499 \\
{\tt hep-th/0307206} \\
July, 2003 \\
\end{flushright}
\vspace{3mm}
\begin{center}
{\Large
{\bf Stability and Fluctuations on Walls
} 
\\
\vspace{2mm}{\bf in ${\cal N}=1$ Supergravity
}
} 
\\[12mm]
\vspace{5mm}

\normalsize

  {\large \bf 
  Minoru~Eto~$^{a}$}
\footnote{\it  e-mail address: 
meto@th.phys.titech.ac.jp
},  
  {\large \bf 
  Nobuhito~Maru~$^{b}$}
\footnote{\it  e-mail address: 
maru@postman.riken.go.jp, Special Postdoctoral Researcher
}  
~and~~ {\large \bf 
Norisuke~Sakai~$^{a}$}
\footnote{\it  e-mail address: 
nsakai@th.phys.titech.ac.jp
}

\vskip 1.5em

{ \it $^{a}$Department of Physics, Tokyo Institute of 
Technology \\
Tokyo 152-8551, JAPAN  \\
and \\
  $^{b}$Theoretical Physics Laboratory \\ 
  RIKEN (The Institute of Physical and Chemical Research) \\
2-1 Hirosawa, Wako, Saitama 351-0198, JAPAN
 }
\vspace{15mm}
{\bf Abstract}\\[5mm]

{\parbox{14cm}{\hspace{5mm}
%%%%%%%%%%%%%%%%%%%%%%%%%%%%%%%%%%%%%%%%%%%%%%%%%%
%%%%%%%%%%
The recently found non-BPS multi-wall 
configurations in the ${\cal N}=1$ supergravity in 
four dimensions is shown to have no tachyonic scalar 
fluctuations 
without 
additional stabilization mechanisms. 
Mass of radion (lightest massive fluctuation) 
is found to be proportional to 
$\Lambda {\rm e}^{-\pi\Lambda R/2}$, where $\Lambda $ is 
the inverse width of the wall and 
 $ R$ is the radius of compactified dimension. 
We obtain localized massless graviton and gravitino 
forming a supermultiplet with respect to 
the Killing spinor. 
The relation between the bulk energy density 
and the boundary energy density (cosmological constants) 
is an automatic consequence of 
the field equation and Einstein equation. 
In the limit of vanishing gravitational coupling, 
the Nambu-Goldstone modes are reproduced. 
%%%%%%%%%%%%%%%%%%%%%%%%%%%%%%%%%%%%%%%%%%%%%%%%%%
%%%%%%%%%%
}}
\end{center}
\vfill
\newpage
\setcounter{page}{1}
\setcounter{footnote}{0}
\renewcommand{\thefootnote}{\arabic{footnote}}

%%%%%%%%%%%%%%%%%%%%%%%%%%%%%%%%%%%%%%
\section{Introduction}

In the brane-world scenario \cite{LED, RS1, RS2}, 
our four-dimensional world is to be realized on 
topological defects such as walls. 
To obtain realistic unified theories beyond the standard 
model, supersymmetry (SUSY) has been most useful 
\cite{DGSW}. 
Moreover, SUSY helps to construct topological defects 
like walls as BPS states \cite{WittenOlive} 
that preserve part of SUSY. 
 For a realistic model, 
understanding SUSY breaking has been an important problem,  
which is addressed in the SUSY brane-world scenario 
extensively \cite{BULK}--\cite{MSSS}. 
Models have been constructed that realize one such idea : 
coexistence of BPS and anti-BPS walls produces 
SUSY breaking automatically \cite{MSSS}. 
In particular, the SUSY breaking effects are suppressed 
exponentially as a function of distance between walls. 
On the other hand, non-BPS multi-wall configurations 
are not protected by SUSY and need not be stable. 
Such non-BPS wall configurations was successfully stabilized 
by introducing topological quantum numbers, 
such as a winding number \cite{MSSS2, SakaiSugisaka}. 
The physical reason behind the stability is simple :  
a BPS wall and an anti-BPS wall with winding numbers 
generally exert repulsion, which then pushes each other 
at anti-podal points of the compactified dimension.  

One of the most attractive models in the brane-world scenario 
is the model with the warped metric \cite{RS1, RS2}. 
A possible solution of the gauge hierarchy problem 
was proposed in the two brane-model \cite{RS1}, and 
a localization of graviton on a single brane was found 
even in a noncompact space 
\cite{RS2} at the cost of fine-tuning between bulk 
cosmological constant and boundary cosmological constant 
at orbifold fixed points. 
Supersymmetrization of the thin-wall model has also been 
constructed in five dimensions \cite{ABN}--\cite{FLP}. 
It is natural to ask if the infinitely thin branes in these 
models can be replaced by physical smooth wall configurations 
made out of scalar fields \cite{CGR}--\cite{SkTo}. 
We have succeeded in constructing BPS as well as non-BPS 
solutions in the ${\cal N}=1$ supergravity coupled with 
a chiral scalar multiplet in four dimensions \cite{EMSS}. 
A similar BPS solution has also been constructed in 
five-dimensional supergravity \cite{AFNS,Eto:2003ut}. 
In the limit of vanishing gravitational coupling 
$\kappa \rightarrow 0$, 
our model reduces to the model having the exact solution of 
non-BPS multi-walls \cite{MSSS2}. 
Therefore the model is likely to be stable thanks 
to the winding number near the weak 
gravity limit. 
However, we need to address the issue of stability in the 
presence of gravity, since the radius of the extra dimension 
is now a dynamical variable which might introduce 
instability into the model. 
There have been a number of works to analyze the stability of 
the infinitely thin wall \cite{GiLa}--\cite{CsabaCsaki}, 
especially in the presence of a stabilizing mechanism 
due to Goldberger and Wise \cite{GoWi}. 

The purpose of our paper is to study the stability of the 
model with winding number in the presence of gravity 
and to analyze 
the mass spectrum of fluctuations on the BPS and non-BPS 
solutions. 
We find that there are zero modes of transverse traceless 
fluctuations localized on the wall which play the role of 
the graviton in our world on the wall. 
The BPS solution has also gravitino zero mode which is 
localized on the wall and forms a supermultiplet 
with the graviton under the surviving supergravity 
transformation with the Killing spinor of the BPS solution. 
We obtain that 
the BPS solution has no other zero modes, 
and no tachyonic fluctuations. 
For instance, we find that possible additional 
massless tensor and scalar modes 
are either gauge degrees of freedom or unphysical 
(the mode function is not normalizable). 
As for the non-BPS solution, we find 
that another possible zero modes of 
the transverse traceless fluctuations of metric 
can be gauged away and that there exists no zero mode 
other than the graviton localized on the wall. 
To obtain a concrete estimate of the mass spectrum, 
we need to use approximations. 
We use small width approximation where the width 
$\Lambda^{-1}$ 
of the wall is small compared to the radius $R$ 
of compactified extra dimension. 
We find that the non-BPS solution 
has no tachyonic 
fluctuations in spite of the 
dynamical role 
played by the radius of the compactified dimension. 
Tensor as well as scalar fluctuations have massive modes, 
without any tachyons. 
This result shows that our non-BPS solution 
is stable without introducing an additional stabilizing 
mechanism such as the Goldberger-Wise mechanism \cite{GoWi}. 

The lightest massive scalar mode is usually called radion. 
We can evaluate the mass of the radion 
on our non-BPS background at least for $R \gg \Lambda^{-1}$, 
where $R$ is the radius of the compactified dimension and 
$\Lambda^{-1}$ is the width of the wall. 
We find that the mass squared of the radion is given by 
\begin{equation}
m^2_0 \propto \Lambda^2 
e^{-\pi R \Lambda} 
\label{eq:radion-mass1}
\end{equation}
It is interesting to note 
that the mass scale is given by the inverse 
wall width $\Lambda$, and that it 
becomes exponentially light 
as a function of the distance $\pi R$ 
between the two walls. 
This behavior is precisely the same as the previous 
model in the global SUSY case \cite{MSSS2}. 

Modes of fermions including gravitino are also analyzed. 
We find that the Nambu-Goldstone modes can be reproduced 
in the limit of vanishing gravitational coupling 
both for bosonic and fermionic modes.

Our BPS solution has a smooth limit of thin walls where 
it reproduces the Randall-Sundrum model \cite{EMSS}. 
In the original Randall-Sundrum model, 
the fine-tuning was necessary between the 
boundary and the bulk cosmological 
constants. 
However, the necessary relation between 
bulk and boundary cosmological constants is now an 
automatic consequence of the equation 
of motion of scalar fields and Einstein equation 
 in our model. 
We no longer need to impose a fine-tuning on input parameters 
of the model. 

Sec.2 summarizes our model and solutions briefly. 
Sec.3 separates various bosonic modes with respect to 
the surviving Lorentz symmetry (tensor and scalar modes) 
and addresses the question of stability of the BPS 
solution. 
Sec.4 discusses the stability of non-BPS solution and 
evaluates the mass of the radion. 
Sec.5 deals with the fermionic modes. 
The gauge fixing to the Newton 
gauge is  justified in Appendix A, and some illustrative 
cases of potential in the conformal coordinate 
are worked out in 
Appendix B.

\section{Brief review of BPS domain wall in SUGRA}
\subsection{Lagrangian and BPS equations
}

We consider a chiral multiplet containing scalar 
$\phi$ and fermion 
$\chi$  with the minimal kinetic term and the superpotential 
$P$, and 
the gravity multiplet 
containing vielbein $e_m{^{\underline{a}}}$ and gravitino 
$\psi_m{^{\alpha}}$. 
The local Lorentz vector indices are denoted by 
letters with the underline as $\underline{a}$, 
and the vector indices 
transforming under general coordinate transformations are 
denoted by Latin letters as $m, n=0, \dots, 3$. 
The left(right)-handed spinor indices\footnote{
We follow conventions of Ref.\cite{WessBagger} for the 
spinor and other notations.} 
are denoted by undotted (dotted) indices as 
${\alpha} ({\dot \alpha})$. 
Then the $\mathcal{N}=1$ supergravity Lagrangian 
is given in 
four-dimensional spacetime as \cite{WessBagger}
\begin{eqnarray}
e^{-1}\mathcal{L} &=&
- \frac{1}{2\kappa^2}R 
+ \varepsilon^{klmn}\bar\psi_k\bar\sigma_l
\tilde{\mathcal{D}}_m\psi_n
\nonumber\\
&&- g^{mn}\partial_m\phi^*\partial_n\phi 
- {\rm e}^{\kappa^2\phi^*\phi}
\left(|D_\phi P|^2 - 3\kappa^2|P|^2\right)
-i \bar\chi\bar\sigma^m\mathcal{D}_m\chi
\nonumber\\
&& - \frac{\sqrt{2}}{2}\kappa
\left(\partial_n\phi^*\chi\sigma^m\bar\sigma^n\psi_m
+ \partial_n\phi\bar\chi\bar\sigma^m\sigma^n
\bar\psi_m\right)
\nonumber\\
&&+ \frac{\kappa^2}{4}
\left(i\varepsilon^{klmn}\psi_k\sigma_l\bar\psi_m 
+ \psi_m\sigma^n\bar\psi^m\right)\chi\sigma_n\bar\chi
- \frac{\kappa^2}{8}\chi\chi\bar\chi\bar\chi
\nonumber\\
&&- {\rm e}^{\frac{\kappa^2}{2}\phi^*\phi}\bigg[
\kappa^2\left(P^*\psi_m\sigma^{mn}\psi_n 
+ P\bar\psi_m\bar\sigma^{mn}\bar\psi_n\right)\nonumber\\
&&+\frac{i\kappa}{\sqrt{2}}
\left(D_\phi P\chi\sigma^m\bar\psi_m 
+ D_{\phi^*}P^*\bar\chi\bar\sigma^m\psi_m\right)
\nonumber\\
&& + \frac{1}{2}
\left(\mathcal{D}_\phi D_\phi P \chi\chi 
+ \mathcal{D}_{\phi^*}D_{\phi^*}P^*\bar\chi\bar\chi\right)
\bigg],
\label{SUGRA_Lag}
\end{eqnarray}
where the gravitational coupling $\kappa$ is 
the inverse of the four-dimensional 
Planck mass $M_{\rm Pl}$,
$g_{mn}$ is the metric of the spacetime and $e$ is 
the determinant of the 
vierbein $e_m{^{\underline{a}}}$. 
The generalized
supergravity covariant derivatives are defined as follows : 
\begin{eqnarray}
\label{covder}
\begin{array}{rll}
\mathcal{D}_m\chi &= &\partial_m\chi + \chi\omega_m 
- \dfrac{i\kappa^2}{2}{\rm Im}\!
\left[\phi^*\partial_m\phi\right]\chi,\\
\tilde{\mathcal{D}}_m\psi_n 
&= &\partial_m\psi_n + \psi_n\omega_m 
+ \dfrac{i\kappa^2}{2}{\rm Im}\!
\left[\phi^*\partial_m\phi\right]\psi_n,\\
D_\phi P &= &\partial_\phi P + \kappa^2\phi^*P,\\
\mathcal{D}_\phi D_\phi P 
&= &\partial_\phi^2P + 2\kappa^2\phi^*D_\phi P 
- \kappa^4\phi^{*2}P,
\end{array}
\end{eqnarray}
where $\omega_m$ is the spin connection and 
we use the notation ${\rm Im}\!\left[X\right]
\equiv \dfrac{X-X^*}{2i}$ in what follows. 
The scalar potential in  the supergravity Lagrangian 
(\ref{SUGRA_Lag}) is given by 
\begin{eqnarray}
V(\phi,\phi^*) = {\rm e}^{\kappa^2\phi^*\phi}
\left(|D_\phi P|^2 - 3\kappa^2|P|^2\right).
\label{eq:scalar-potential}
\end{eqnarray}

The above Lagrangian
(\ref{SUGRA_Lag}) is invariant under the 
supergravity transformation : 
\begin{eqnarray}
%\label{SUGRAtrf}
\begin{array}{rll}
\delta_\zeta e_m{^{\underline{a}}} 
&= &i\kappa\left(\zeta\sigma^{\underline{a}}\bar\psi_m 
+ \bar\zeta\bar\sigma^{\underline{a}}\psi_m\right),\\
\delta_\zeta \psi_m 
&= &2\kappa^{-1}\mathcal{D}_m\zeta 
+ i\kappa{\rm e}^{\frac{\kappa^2}{2}\phi^*\phi}
P\sigma_m\bar\zeta
- \dfrac{i\kappa}{2}\sigma_{mn}\zeta\chi\sigma^n\bar\chi
- \dfrac{i\kappa^2}{2}{\rm Im}\!
\left[\phi^*\delta_\zeta\phi\right]\psi_m,\\
\delta_\zeta \phi &= &\sqrt{2}\ \zeta\chi,\\
\delta_\zeta \chi &= &i\sqrt{2}\ \sigma^m\bar\zeta
\hat{\mathcal{D}}_m\phi 
- \sqrt{2}\ {\rm e}^{\frac{\kappa^2}{2}\phi^*\phi}
D_{\phi^*}P^*\zeta
+ \dfrac{i\kappa^2}{2}{\rm Im}\!
\left[\phi^*\delta_\zeta\phi\right]\chi,
\end{array}
\label{eq:SUGRAtransf}
\end{eqnarray}
where $\zeta$ is a local SUSY transformation 
parameter and the covariant
derivatives are given by 
\begin{eqnarray}
\begin{array}{rll}
\hat{\mathcal{D}}_m\phi 
&= &\partial_m\phi - \dfrac{\sqrt{2}}{2}
\kappa\bar\psi_m\bar\chi,\\
\mathcal{D}_m\zeta &= &\partial_m\zeta + \zeta\omega_m 
+ \dfrac{i\kappa^2}{2}{\rm Im}\!
\left[\phi^*\partial_m\phi\right]\zeta.
\end{array}
\end{eqnarray}

Next we turn to derive the equations of motion 
for solutions which
depend on only one ``extra'' coordinate $x^2=y$ under 
the warped metric Ansatz
\begin{eqnarray}
ds^2 = g_{mn}dx^mdx^n 
= {\rm e}^{2A(y)}\eta_{\mu\nu}dx^\mu dx^\nu + dy^2 
\quad (\mu,\nu = 0,1,3),\label{warped_metric}
\end{eqnarray}
where Greek indices $\mu=0, 1, 3$ denote three-dimensional 
vector transforming under general coordinate 
transformations, 
and 
$\eta_{\mu\nu} = {\rm diag}(-,+,+)$ denotes three 
dimensional flat
spacetime metric. 
All the geometrical quantities can be written 
in terms of the function $A(y)$ in the 
warp factor and its derivatives with respect to 
the extra coordinate
$y$. 
For later convenience, we write formulas in general 
$D$ space-time dimensions in the following : 
\begin{enumerate}
\item vierbein
\begin{eqnarray}
e_m{^{\underline{a}}} 
= {\rm diag}\left({\rm e}^A,\ {\rm e}^A,\ 1,\ 
{\rm e}^A\right),\quad
e_{\underline{a}}{^m} 
= {\rm diag}\left({\rm e}^{-A},\ 
{\rm e}^{-A},\ 1,\ {\rm e}^{-A}\right),
\end{eqnarray}
\item spin connection
\begin{eqnarray}
(\chi \omega_m)_\alpha = {1 \over 2}
\omega_{m\underline{ab}}
\left(\sigma^{\underline{ab}}\right)_\alpha{}^\beta 
\chi_\beta, 
\qquad 
\omega_{m\underline{ab}} = 
\dot{A}\left(\delta_{\underline{a}}{^2}e_{\underline{b}m}
- \delta_{\underline{b}}{^2}e_{\underline{a}m}\right),
\label{eq:spin-connec}
\end{eqnarray}
\item Ricci tensor
\begin{eqnarray}
R_{mn} = {\rm e}^{2A}\left(
\ddot{A} + (D-1)
\dot{A}^2\right)
\eta_{\mu\nu}\delta_m{^\mu}\delta_n{^\nu}
+ (D-1)
\left(
\ddot{A} + 
\dot{A}^2\right)\delta_m{^2}\delta_n{^2},
\end{eqnarray}
\end{enumerate}
where a dot denotes a derivative with respect to 
$y$, $\dot A\equiv dA/dy$, 
and we turn off all the fermionic fields as a tree level
solution. The energy momentum tensor is given in terms 
of the scalar potential $V(\phi, \phi^*)$ 
in (\ref{eq:scalar-potential})
\begin{eqnarray}
T_{mn} = \partial_m\phi^*\partial_n\phi 
+ \partial_m\phi\partial_n\phi^* 
- g_{mn}\left(g^{kl}\partial_k\phi^*\partial_l\phi 
+ V(\phi,\phi^*)\right).
\end{eqnarray}

Plugging these into the Einstein 
equation\footnote{We define 
$\tilde{T}_{mn}\equiv 
T_{mn} - \frac{1}{D-2}g_{mn}T^k{_k}$. 
}
 $R_{mn} = - \kappa^2\tilde{T}_{mn}$, 
we obtain 
\begin{eqnarray}
\ddot{A} = -{2\over D-2}\kappa^2
\dot\phi^*\dot\phi,\quad
\dot{A}^2 = \frac{2\kappa^2}{(D-1)(D-2)}\left(
\dot\phi^*\dot\phi-V(\phi,\phi^*)\right).
\label{Einstein_eq}
\end{eqnarray}
The field equation for the scalar $\phi$ 
in the chiral multiplet 
takes the form : 
\begin{eqnarray}
\ddot\phi
+ (D-1) \dot{A}\dot\phi
= \frac{\partial V}{\partial\phi^*}.
\label{field_eq}
\end{eqnarray}
Notice that only two out of the three equations in 
Eqs.(\ref{Einstein_eq}) 
and (\ref{field_eq}) are independent 
(assuming only one real component, say the real part 
of the scalar field $\phi$ is nontrivial in the solution). 
Any one of three equations are automatically satisfied 
if others are satisfied.

It is well known that special type of solutions 
for these nonlinear
second order differential equations are 
obtained as solutions of a set 
of the first order differential equations, 
the so-called BPS equations 
which guarantees the partial conservation of SUSY. 
Similarly to the global SUSY case, the BPS 
equations can be derived from
the half SUSY condition where we parametrize 
the conserved SUSY
parameter as 
\begin{eqnarray}
\zeta(y) = {\rm e}^{i\theta(y)}
\sigma^{\underline{2}}\bar\zeta(y).
\label{eq:half-susy}
\end{eqnarray} 
That is, we demand
that the bosonic configuration should satisfy 
$\delta_\zeta\chi =
\delta_\zeta\psi_m=0$ for the parameter 
$\zeta(y)$ in Eq.(\ref{eq:half-susy}). 
The BPS equations
for the metric are derived from the condition for 
the gravitino. 
 From $m=\mu=0, 1, 3$ components the first order 
 equation for the warp
factor $A$ is derived : 
\begin{eqnarray}
0 = \delta_\zeta\psi_\mu 
= \kappa^{-1}{\rm e}^A
\left[{\rm e}^{i\theta}\dot{A} 
+ i\kappa^2{\rm e}^{\frac{\kappa^2}{2}\phi^*\phi}
P\right]\sigma_{\underline{\mu}}\bar\zeta ,
\end{eqnarray}
\begin{eqnarray}
\dot{A} 
= - i\kappa^2{\rm e}^{-i\theta}
{\rm e}^{\frac{\kappa^2}{2}\phi^*\phi}P.\label{BPS_A}
\end{eqnarray}
From $m=2$ component we find the first order 
equation for the
Killing spinor $\zeta$ : 
\begin{eqnarray}
0 = \delta_\zeta\psi_2 = 2\kappa^{-1}
\left[\dot\zeta 
+ \frac{i\kappa^2}{2}{\rm Im}\!
\left[\phi^*\dot\phi
\right]\zeta
+ \frac{i\kappa^2}{2}{\rm e}^{-i\theta}
{\rm e}^{\frac{\kappa^2}{2}\phi^*\phi}P\zeta\right].
\end{eqnarray}
Rewriting the half SUSY condition as 
$\zeta_{\underline{\alpha}} ={\rm
e}^{\frac{i}{2}\left(\theta + \frac{\pi}{2}\right)}
|\zeta_{\underline{\alpha}}|$ and substituting
it into the above equation, we find \cite{CGR}
\begin{eqnarray}
\dot{|\zeta_{\underline{\alpha}}|} 
= \frac{\dot{A}}{2}
|\zeta_{\underline{\alpha}}|,\quad
\dot\theta 
= - \kappa^2{\rm Im}\!\left[\phi^*\dot\phi
\right].\label{BPS_kill}
\end{eqnarray}
On the other hand, the first order equation 
for the matter field $\phi$
is derived from the half SUSY condition for 
the matter fermion $\chi$ : 
\begin{eqnarray}
\dot\phi 
= -i{\rm e}^{i\theta}{\rm e}^{\frac{\kappa^2}{2}
\phi^*\phi}D_{\phi^*}P^*.\label{BPS_phi}
\end{eqnarray}
Eq.(\ref{BPS_A}), (\ref{BPS_kill}) and 
(\ref{BPS_phi}) are collectively
called BPS equations. 
One can easily show that solutions of the BPS
equations satisfy the equations of motion 
(\ref{Einstein_eq}) and
(\ref{field_eq}).
Notice that the Eq.(\ref{BPS_phi}) and the 
second equation of
Eq.(\ref{BPS_kill}) do not contain the metric, so we
can solve this as if the scalar field decouples 
from gravity.
Once the configuration of the scalar field 
$\phi$ and the phase $\theta$
are determined, the
warp factor $A$ is obtained from Eq.(\ref{BPS_A}). 
Finally, the Killing
spinor $\zeta$ is also determined from the first equation of
Eq.(\ref{BPS_kill}).

\subsection{Exact BPS solution}
Recently, we found the exact BPS solutions 
for the periodic model in
SUGRA \cite{EMSS}, by allowing the gravitational 
correction for the superpotential as
follows
\begin{eqnarray}
P(\phi) 
= {\rm e}^{-\frac{\kappa^2}{2}\phi^2}
\times \frac{\Lambda^3}{g^2}
\sin\frac{g}{\Lambda}\phi,\label{P_mod}
\end{eqnarray}
where $\Lambda$ is a coupling with unit mass 
dimension and $g$ is a
dimensionless coupling. 
We introduced this modification 
for the superpotential in
SUGRA to maintain the periodicity of the model 
with the aid of the
K\"ahler transformation.
This modification for the superpotential gives 
SUSY vacua which do not 
depend on the gravitational coupling $\kappa$. 
This was crucial for us 
to obtain the exact BPS
solutions in SUGRA.
The superpotential (\ref{P_mod}) yields the 
following scalar
potential : 
\begin{eqnarray}
V = \frac{\Lambda^4}{g^2}
{\rm e}^{2\kappa^2({\rm Im}\left[\phi\right])^2}
\left[
\left|\cos\frac{g}{\Lambda}\phi 
- \frac{2i\kappa^2\Lambda}{g}{\rm Im}\!\left[\phi\right]
\ \sin\frac{g}{\Lambda}\phi\right|^2
- \frac{3\kappa^2\Lambda^2}{g^2}
\left|\sin\frac{g}{\Lambda}\phi\right|^2
\right].
\end{eqnarray}
The SUSY vacua are determined from the condition 
$D_\phi P = 0$. For the
above modified superpotential we find that 
the SUSY vacua 
are periodically 
distributed at $\phi =
\dfrac{\Lambda}{g}\left(\dfrac{\pi}{2} 
+ n\pi\right),\ (n\in\mathbb{Z})$
on the real axis in the complex $\phi$ plane. 

In order to determine the scalar field configuration, 
we need to solve
the second equation of Eq.(\ref{BPS_kill}) 
for $\theta$ together with
the equation for scalar field : 
\begin{eqnarray}
\dot\phi 
= - i{\rm e}^{i\theta}{\rm e}^{i\kappa^2\phi^*{\rm Im}[\phi]}
\frac{\Lambda^2}{g}
\left[\cos\frac{g}{\Lambda}\phi^* 
+ \frac{2i\Lambda\kappa^2}{g}{\rm Im}[\phi]
\sin\frac{g}{\Lambda}\phi^*\right].
\label{eq:matterBPSeq}
\end{eqnarray}
To solve Eqs.(\ref{BPS_kill}) and 
(\ref{eq:matterBPSeq}), 
we choose $\phi_{\rm I} \equiv {\rm Im}[\phi] = 0$ and 
$\theta = \pm\dfrac{\pi}{2}$
at a point, say $y=y_i$ as an initial condition 
for the imaginary part $\phi_{\rm I}(y)$ of the scalar
field and the phase $\theta(y)$. 
Then these equations  
tell that
$\dot\phi_{\rm I} 
= \dot\theta 
= 0$ at $y=y_i$. Therefore we 
find 
$\phi_{\rm I} = 0$ and $\theta = \pm\dfrac{\pi}{2}$ at any
$y$. 
At this stage, only the real part of $\phi$ has a nontrivial 
configuration in the extra dimension $y$. 
We shall call those scalar fields that have 
nontrivial configurations as a function of the 
coordinate of extra dimension, 
as ``active'' scalar fields.
The scalar potential along $\phi_{\rm I}=0$ surface is given 
by the 
following potential $V_{\rm R}$ for the real part 
$\phi_{\rm R} \equiv\dfrac{\phi+\phi^*}{2}$ of 
the scalar field : 
\begin{eqnarray}
V_{\rm R}(\phi_{\rm R}) 
= \frac{\Lambda^4}{g^2}
\left[\cos^2\frac{g}{\Lambda}\phi_{\rm R}
- \frac{3\kappa^2\Lambda^2}{g^2}
\sin^2\frac{g}{\Lambda}\phi_{\rm R}\right].
\label{eq:real-scalar-pot}
\end{eqnarray}
It has been shown that 
the following form of scalar potential 
with a real ``superpotential'' $\hat P(\phi_{\rm R})$ 
of a real scalar field $\phi_{\rm R}$ 
ensures the existence of a stable AdS vacuum 
in gravity theories in $D$ dimensions 
\cite{Boucher, PKT} :  
\begin{eqnarray}
V_{\rm R} = {D-2 \over 2}\left[{D-2 \over 2}
\left(\frac{d\hat P}{d\phi_{\rm R}}\right)^2
-(D-1)\kappa^2\hat P^2\right],\label{potential_ads}
\end{eqnarray}
 if there is a critical point in $\hat P(\phi_R)$, 
 even though supersymmetry is not required in this form. 
Let us note that our scalar potential is compatible with 
the above form of the scalar potential. 
In our case, the ``superpotential'' is given by 
\begin{eqnarray}
\hat P(\phi_{\rm R}) = \frac{\Lambda^3}{g^2}\sin
\frac{g}{\Lambda}\phi_{\rm R} .
\end{eqnarray}
Since this $\hat P$ has critical points at
$\phi_R=\dfrac{\Lambda}{g}
\left(\dfrac{\pi}{2}+n\pi\right)$,
our scalar potential 
$V_{\rm R}(\phi_{\rm R})$ in 
Eq.(\ref{eq:real-scalar-pot})
has these critical points as 
stable AdS vacua.

The remaining BPS equations for the active scalar
field and the warp factor are of the form : 
\begin{eqnarray}
\dot\phi_{\rm R} 
= \pm \frac{d\hat P}{d\phi_{\rm R}}
= \pm \frac{\Lambda^2}{g}
\cos\frac{g}{\Lambda}\phi_{\rm R},\quad
\dot{A} 
= \mp \kappa^2\hat P
= \mp \frac{\kappa^2\Lambda^3}{g^2}
\sin\frac{g}{\Lambda}\phi_{\rm R}.
\label{BPS_eq_R}
\end{eqnarray}
Let us solve these BPS equations 
by choosing a SUSY vacuum
$\phi_{\rm R}
=\dfrac{\Lambda}{g}\left(\mp(-1)^n\dfrac{\pi}{2}+n\pi\right)$ 
as an initial 
condition at $y=-\infty$. 
We shall consider the solution for the 
 BPS equations (\ref{BPS_eq_R}) with the sign 
correlated to the sign of the initial condition at 
$y=-\infty$. 
The exact BPS solutions are found to be of the form : 
\begin{eqnarray}
\phi_{\rm R} 
= \frac{\Lambda}{g}\left[(-1)^n
\left\{2\tan^{-1}{\rm e}^{\pm\Lambda(y-y_0)} 
- \frac{\pi}{2}\right\} + n\pi\right],\quad
{\rm e}^{A} 
= \left[\cosh\Lambda(y-y_0)
\right]^{-\frac{k}{\Lambda}},
\label{BPS_sol}
\end{eqnarray}
where $k\equiv \dfrac{\kappa^2\Lambda^3}{g^2}$ 
is the inverse of the
curvature radius of the AdS spacetime at infinity.
These solutions interpolate between the two SUSY vacua, 
from $\phi_{\rm R} =
\dfrac{\Lambda}{g}\left(\mp(-1)^n \dfrac{\pi}{2} 
+ n\pi\right)$ at $y =
-\infty$ to 
$\phi_{\rm R} = \dfrac{\Lambda}{g}\left(\pm(-1)^n\dfrac{\pi}{2} +
n\pi\right)$ at $y = +\infty$. 
We denote  $y_0$ the modulus parameter of these
solutions  and we suppress an integration constant 
for $A$ which amounts to
an irrelevant normalization constant of metric.
Eq.(\ref{BPS_kill})  determines the 
Killing spinors which has two real Grassmann parameters 
$\epsilon_1, \epsilon_2$ 
corresponding to the two conserved SUSY directions 
on the BPS solution\footnote{
These Killing spinors are the corrected results of those 
in our previous work \cite{EMSS}. 
} : 
\begin{eqnarray}
\zeta 
= {\rm e}^{\frac{i}{2}\left(\theta 
+ \frac{\pi}{2}\right)}
{\rm e}^{\frac{A}{2}}\times
\left(\begin{array}{c}
 \epsilon_1 \\ 
 \epsilon_2 
   \end{array}\right), 
\label{eq:Killingspinor}
\end{eqnarray}
\begin{eqnarray}
{\rm e}^{\frac{i}{2}\left(\theta 
+ \frac{\pi}{2}\right)}
{\rm e}^{\frac{A}{2}}
 = 
\left\{
\begin{array}{ll}
i\left[\cosh\Lambda(y-y_0)\right]^{
-\frac{k}{2\Lambda}},
\quad&{\rm for}\ \theta = \dfrac{\pi}{2},\\
\left[\cosh\Lambda(y-y_0)\right]^{
-\frac{k}{2\Lambda}},
\quad&{\rm for}\ \theta = -\dfrac{\pi}{2}.
\end{array}
\right.
\end{eqnarray}

Our model has a smooth limit of thin walls where 
it reproduces the Randall-Sundrum model \cite{EMSS}. 
Notice that we do not need any fine-tuning of 
input parameters of the model, in contrast to the original 
Randall-Sundrum model. 
The necessary fine-tuning between 
bulk and boundary cosmological constants is now an 
automatic consequence of the equation 
of motion of scalar fields and Einstein equation 
 in our model.

\subsection{non-BPS solution}

Assuming that only single real scalar field $\phi_{\rm R}$ 
has nontrivial 
classical configuration, 
the equations (\ref{Einstein_eq}) and (\ref{field_eq}) 
reduce to
\begin{eqnarray}
\ddot{A}
=-{2\kappa^2 \over D-2}\dot\phi_{\rm R}^2
,\qquad 
\dot{A}^2 
= \frac{2\kappa^2}{(D-1)(D-2)}
\left(\dot\phi_{\rm R}^2
-V_{\rm R}\right),
\qquad 
\ddot\phi_{\rm R}
+ (D-1) \dot{A}\dot\phi_{\rm R}
= \frac{1}{2}\frac{dV_{\rm R}}{d\phi_{\rm R}}.
\label{eom_real_phi}
\end{eqnarray}
It has been shown that the above set 
of coupled second order differential equations is 
equivalent to the following set of nonlinear 
differential equations \cite{DFGK, SkTo}. 
Given the scalar potential $V_{\rm R}(\phi_{\rm R})$, 
we should find a real function 
$W(\phi_{\rm R})$ by solving the following first order 
nonlinear differential 
equation 
\begin{equation}
{d W(\phi_{\rm R}) \over d\phi_{\rm R}}
= \pm {2 \over D-2}
\sqrt{V_{\rm R}(\phi_{\rm R})+
{(D-1)(D-2) \over 2}\kappa^2W^2(\phi_{\rm R})} .
\label{eq:nonlinear-eq}
\end{equation}
Then $\phi_{\rm R}(y)$ and $A(y)$ are obtained 
by solving the 
following two first order differential equations 
\begin{equation}
\dot\phi_{\rm R}(y)=
{D-2 \over 2}\dfrac{d W(\phi_{\rm R})}{d \phi_{\rm R}}, 
\qquad 
\dot A(y) = - \kappa^2W(\phi_{\rm R}) .
\label{phi-A}
\end{equation}
If we choose the ``superpotential'' $\hat P$ as a real 
function $W$, (\ref{eq:nonlinear-eq}) and 
 (\ref{phi-A}) are satisfied by the 
 scalar potential (\ref{potential_ads}) and 
the BPS equations (\ref{BPS_eq_R}). 
Therefore these set of first order nonlinear 
differential equations includes all the BPS solutions as 
part of the solutions. 
However, it is important to realize that 
(\ref{eq:nonlinear-eq}) and 
(\ref{phi-A}) are equivalent to the set of Einstein 
equation and the scalar filed equation, and hence 
give all the non-BPS solutions as well. 

We have been able to construct non-BPS multi-wall 
solutions to 
the Einstein equation (\ref{Einstein_eq}) 
and the field equation 
(\ref{field_eq}) using the above method of 
nonlinear equations \cite{EMSS}. 
We have also found that BPS solutions are the only 
solution that do not encounter singularities 
at any finite $y$. 
To obtain any other regular solution, especially non-BPS 
solutions, 
negative cosmological constant has to be introduced 
at some boundary. 
Since we are interested in periodic array of walls 
where extra dimension can be identified as a torus 
$S^1$ with possible division by discrete groups 
(orbifolds), we introduced 
the cosmological constant and obtained a number of 
interesting non-BPS solutions \cite{EMSS}. 

The above nonlinear differential equation 
(\ref{eq:nonlinear-eq}) gives a set of solution curves 
which fill once and only once 
the entire $(\phi_{\rm R}, W)$ plane except 
forbidden regions defined by 
$V_{\rm R}+(D-1)(D-2)\kappa^2W^2/2 \le 0$. 
Let us denote the solution curve starting from an 
initial condition $W_0$ at $\phi_{\rm R} = \phi_{{\rm R},0}$ as 
$W(\phi_{\rm R}; (\phi_{{\rm R},0}, W_0))$. 
A boundary cosmological constant $\lambda_i$ 
at $y_i$ gives a jump of derivative of the function 
$A(y)$ in the warp factor. 
Let us denote the value of the scalar field at the 
boundary $y_i$ as $\phi_{{\rm R},i}$. 
Eq.(\ref{phi-A}) shows that 
this jump of $A(y)$ is satisfied by cutting the 
solution curve and jump to another solution curve 
 at $\phi_{{\rm R},i}$ with the constraint 
\begin{equation}
\lambda_i
=2 \left(W (\phi_{{\rm R},i}+\epsilon)-W (\phi_{{\rm R},i}-\epsilon)\right) .
\label{eq:cosm_const_W}
\end{equation}
Since we are interested in minimum amount of 
inputs at boundaries, we wish to implement 
only the boundary cosmological constant 
without any boundary potential for scalar fields 
$\phi_{\rm R}$, contrary to many other approaches 
characteristic of the Goldberger-Wise type of 
the stabilization mechanism 
\cite{GoWi}, \cite{DFGK}, \cite{SkTo}.  
Therefore we need to maintain the derivative $dW/d\phi_{\rm R}$ 
to be smoothly connected at the boundary. 

Since Eq.(\ref{eq:nonlinear-eq}) gives the same 
value of derivative $dW/d\phi_{\rm R}$ for $\pm W$, 
we can connect the solution curve at any value 
of $\phi_{{\rm R},i}$ if we switch from a solution curve 
going through $W, \phi_{{\rm R},i}$ 
to another one going through $-W, \phi_{{\rm R},i}$. 
Eq.(\ref{eq:cosm_const_W}) gives 
the necessary cosmological constant at this boundary 
as $\lambda=4W(\phi_{{\rm R},i})$. 
There may be other possibilities to connect 
the solution curves, but this is the simplest 
possibility that covers many interesting situations. 

 To be definite, 
we shall consider walls that have simple symmetry property 
under the parity $Z_2$ : $\phi_{\rm R} \rightarrow -\phi_{\rm R}$. 
Let us start a solution curve going through 
$\phi_{\rm R}=0, W_0>0$. 
Then the solution curve goes above the forbidden region. 
To obtain a non-BPS solution which is odd under the 
$Z_2$ transformation, 
we place a boundary at $\phi_{\rm R}=0$ 
with a positive cosmological constant by an amount 
$\lambda_0=4W_0>0$. 
On the other hand, we can place a boundary 
at any $\phi_{\rm R}>0$ with a negative cosmological constant 
$\lambda_1=-4W(\phi_{{\rm R},1}, (\phi_{\rm R}=0, W_0))$. 
However, we 
can obtain a multi-wall solution that have simple 
transformation property under the $Z_2$ by placing 
another boundary at integer multiple of 
$\phi_{\rm R}=\Lambda \pi/(2g)$.

%%%%%%%%%%%%%%%%%%%%%%%%%%%%%%%%%%%%%%%%%%%%%%%%%%%%%%%%%%%%%%%%%%
\begin{figure}[htb]
\begin{center}
\includegraphics[width=15cm
]
{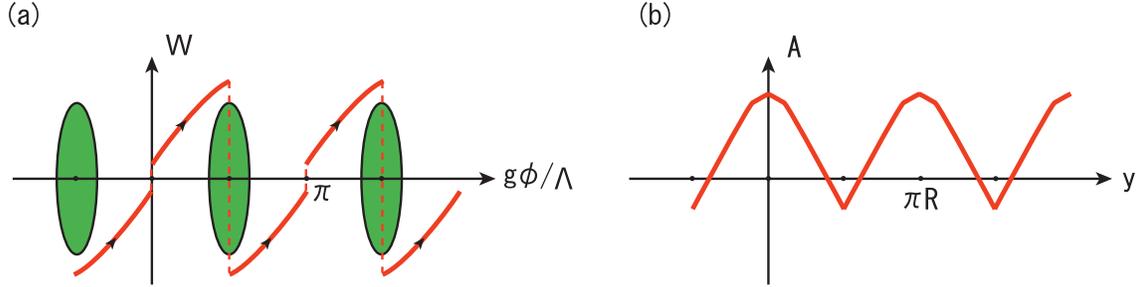}
 \caption{\small 
 Non-BPS solution with unit winding number with 
 the $S^1/(Z_2\times Z_2)$ symmetry. 
 Discontinuities in $ W( \phi)$ at 
 $g\phi/\Lambda=0, \pi$ ($\pi/2, 3\pi/2$) 
 correspond to positive (negative) 
 boundary cosmological  constants. 
 (a) $ \phi  W$ 
 plane, and (b)The function $A(y)$ in 
the warp factor. 
 }
 \label{fig:half_wind_sol}
\end{center}
\end{figure}
%%%%%%%%%%%%%%%%%%%%%%%%%%%%%%%%%%%%%%%%%%%%%%%%%%%%%%%%%%%%%%%%%%
If we place the first boundary at 
the vacuum point $\phi_{\rm R}=\Lambda \pi/(2g)$, 
we obtain a simplest model in the sense 
that the energy density at the 
second boundary at  $\phi_{\rm R}=\Lambda \pi/(2g)$ 
is purely made of negative cosmological constant 
\begin{equation}
\lambda = 
-4W(\phi_{\rm R}=\Lambda \pi/(2g)). 
\label{eq:negaive-cosm-pi/2}
\end{equation}
The magnitude of this negative cosmological constant 
becomes the same as the total energy of the wall 
centered at $\phi_{\rm R}=0$ in the limit of large separation 
of two boundaries. 
Since the solution admits $S^1/(Z_2\times Z_2)$ 
symmetry, we call 
the coordinate at the second boundary $y=\pi R/2$. 
The behavior of this non-BPS solutions in the $W, \phi$ plane 
is illustrated in Fig.\ref{fig:half_wind_sol}(a). 
The corresponding function $A(y)$ in 
the warp factor is illustrated in 
Fig.\ref{fig:half_wind_sol}(b), where 
one should note that $A(y)$ is linear near the 
second boundary 
at $y=\pi R/2$, showing that only the boundary cosmological 
constant exists apart from the bulk cosmological constant 
there.

%%%%%%%%%%%%%%%%%%%%%%%%%%%%%%%%%%%%%%%%%%%%%%%%%%%%%%%%%%%%%%%%%%
\begin{figure}[hbt]
\begin{center}
\includegraphics[width=15cm
]
{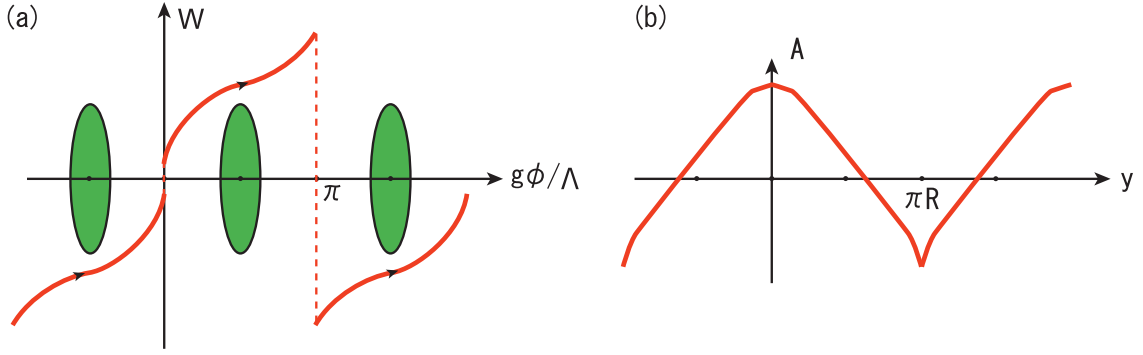}
 \caption{\small 
 Non-BPS solution with unit winding number with 
 the $S^1/Z_2$ symmetry. 
 Discontinuities in $ W( \phi)$ at 
 $g\phi/\Lambda=0$ ($\pi$) 
 correspond to positive (negative) 
 boundary cosmological  constants. 
 (a) $ \phi  W$ 
 plane, and (b)The function $A(y)$ in 
the warp factor. 
 }
 \label{fig:wind_sol}
\end{center}
\end{figure}
%%%%%%%%%%%%%%%%%%%%%%%%%%%%%%%%%%%%%%%%%%%%%%%%%%%%%%%%%%%%%%%%%%
As another solution, we can place the second boundary 
at $\phi_{\rm R}=\Lambda \pi/g$, where the active scalar field 
$\phi_{\rm R}$ develops another wall configuration. 
In this case, the negative cosmological constant 
$-4W(\phi_{\rm R}=\Lambda \pi/g)$ placed at the second 
boundary has magnitude which becomes 
twice  the total energy of the wall 
centered at $\phi_{\rm R}=0$ in the limit of large separation 
of two boundaries. 
The behavior of this non-BPS solutions in the $W, \phi$ plane 
is illustrated  in Fig.\ref{fig:wind_sol}(a). 
The corresponding function $A(y)$ in 
the warp factor is illustrated in 
Fig.\ref{fig:wind_sol}(b), where 
one should note that 
the function $A(y)$ has additional kink behavior 
deviating from the linear exponent 
near the second boundary 
at $y=\pi R$, showing that there is an additional 
smooth positive energy density centered around 
the boundary besides the negative 
boundary cosmological constant in contrast to 
the previous $S^1/(Z_2\times Z_2)$ example in 
Fig.\ref{fig:half_wind_sol}.

\section{Bosonic Fluctuation and 
the BPS Solution}
A  Bogomolo'nyi bound has been derived for 
the  energy density of 
the BPS domain walls in $\mathcal{N}=1$ SUGRA in 
four-dimensional spacetime \cite{CGR}. 
 They used the generalized Israel-Nester-Witten tensor, 
which was originally applied to a simple proof 
of the positive ADM mass conjecture in general relativity. 
However, the ADM mass may not be well-defined for domain 
walls, since they are extended to infinity. 
Therefore it is presumably still useful to check 
that there is really stability of the fluctuation on 
our wall configuration even in the case of BPS 
solutions. 
We shall present a general formalism to analyze the modes 
and their stability, and then apply it to 
the fluctuations around the 
BPS background configurations in this section. 
The equations and 
procedures obtained in 
this section can also be used to the non-BPS 
background solutions with appropriate additional inputs, 
which is dealt with in 
Sec.\ref{sc:stability=nonBPS}.

\subsection{Mode equations for the bosonic sector}

We start with the metric perturbation in the Newton 
gauge \cite{TanakaMontes}, \cite{CsabaCsaki} : 
\begin{eqnarray}
ds^2 = {\rm e}^{2A}\left(\eta_{\mu\nu}+h^{\rm TT}_{\mu\nu} 
+ 2B\eta_{\mu\nu}\right)dx^\mu dx^\nu
+ (1-2(D-3)B)dy^2,
\label{eq:newton-gauge}
\end{eqnarray}
where $h^{\rm TT}_{\mu\nu}$ is transverse traceless 
$\eta^{\mu\nu}h^{\rm TT}_{\mu\nu}=0, \; 
\partial^\mu h^{\rm TT}_{\mu\nu}=0$. 
Some details for the procedure of this gauge 
fixing are given in Appendix A.
This gauge is useful since the linearized equations become 
very simple. 
The linearized Einstein equations in $D$ space-time 
dimensions ($D=4$ in our specific model) read : 
\begin{alignat}{2}
&\left({\rm e}^{-2A}\square_{D-1} +
\partial_y^2
+ (D-1)\dot{A}\partial_y
\right)h^{\rm TT}_{\mu\nu} 
= 0,\label{Newton_1}\\
&\left({\rm e}^{-2A}\square_{D-1} + \partial_y^2
+ (3D-5)\dot{A}\partial_y
+ 2(D-3)\left((D-1)\dot{A}^2
+ 
\ddot{A}\right)
\right)B
= -{2\kappa^2\over D-2}\frac{dV_{\rm R}}{d\phi_{\rm R}}
\varphi_{\rm R},
\label{Newton_2}\\
&\left(\partial_y + (D-3)\dot{A}
\right)B = - {2\kappa^2\over D-2}\kappa^2\dot\phi_{\rm R}
\varphi_{\rm R},\label{Newton_3}
\end{alignat}
where the first line comes from the traceless part 
of $(\mu,\nu)$ component of the linearized 
Einstein equations, the second line from the 
trace part of $(\mu,\nu)$ and
the last from $(\mu,2)$ component. 
The $(2,2)$ component of the linearized
equation is not shown, since it can be derived 
from Eqs.(\ref{Newton_1})-(\ref{Newton_3}). 
The linearized field equations give : 
\begin{alignat}{2}
&\left({\rm e}^{-2A}\square_{D-1} + \partial_y^2
+ (D-1)\dot{A}\partial_y
- \frac{1}{2}\frac{d^2V_{\rm R}}{d\phi_{\rm R}^2}
\right)\varphi_{\rm R}
= -2(D-2) \dot\phi_{\rm R}\partial_yB
- (D-3)\frac{dV_{\rm R}}{d\phi_{\rm R}}B,\label{active}\\
&\left({\rm e}^{-2A}\square_{D-1} + \partial_y^2
+ (D-1)\dot{A}\partial_y
- \frac{1}{2}\frac{\partial^2V}{\partial \phi_{\rm I}^2}
\bigg|_{\rm background}\right)\varphi_{\rm I}
= 0,\label{inert}
\end{alignat}
where $\varphi_{\rm R(I)}$ denotes the real (imaginary) 
part of the fluctuation of the scalar field 
$\phi=\phi_{\rm R}+\varphi_{\rm R}+i\varphi_{\rm I}$ 
around the background field configuration 
$\phi_{\rm R}$.
Notice that the solutions of the linearized 
Einstein equation 
automatically satisfy the 
linearized field equations 
for the active scalar field $\varphi_{\rm R}$. 
Therefore, the Eqs.(\ref{Newton_1})--
(\ref{inert}) constitute the 
full set of independent linearized 
equations for the fields 
$h^{\rm TT}_{\mu\nu},\ B,\ \varphi_{\rm R}$ and
$\varphi_{\rm I}$.

\subsection{Tensor perturbation 
: localized massless graviton}
\label{sc:tensot-perturb}
First we show that the linearized equation 
for the transverse traceless mode (graviton) given in 
Eq.(\ref{Newton_1}) can be brought into a 
Schr\"odinger form. 
It can again be rewritten into a form of the 
supersymmetric quantum mechanics (SQM) which 
ensures the stability of the system.
For that purpose  we change the 
coordinate $y$ into the 
conformally flat coordinate $z$ 
defined as 
\begin{eqnarray}
dz\equiv{\rm e}^{-A(y)}dy, \qquad 
ds^2={\rm e}^{2A(y)}\left(
\eta_{\mu\nu}dx^\mu dx^\nu + dz^2 
\right). 
\end{eqnarray}
We also redefine the field as
$\tilde{h}^{\rm TT}_{\mu\nu} \equiv 
{\rm e}^{{D-2\over 2}A}h^{\rm TT}_{\mu\nu}$. 
In the following we use prime to denote a derivative 
in terms of $z$. 
Then the linearized equation (\ref{Newton_1}) becomes 
\begin{eqnarray}
\square_{D-1}\tilde{h}^{\rm TT}_{\mu\nu}(x,z) 
= \left[-\partial_z^2 + \mathcal{V}_t(z)\right]
\tilde{h}^{\rm TT}_{\mu\nu}(x,z),
\qquad 
\mathcal{V}_t(z) = \left({D-2\over 2}\right)^2A'{^2} 
+ {D-2\over 2}A'', 
\label{eq:schrod-tt}
\end{eqnarray}
where $\mathcal{V}_t(z)$ is the potential 
in this  ``Schr\"odinger'' type equation. 
For our BPS background solution (\ref{BPS_sol}) 
 the Schr\"odinger
potential takes the form : 
\begin{eqnarray}
\mathcal{V}_t(y) = 
\big[\cosh\Lambda(y-y_0)\big]^{-\frac{2k}{\Lambda}}
\left[-\frac{k\Lambda}{\cosh^2\Lambda(y-y_0)}
+ 2k^2\tanh^2\Lambda(y-y_0)\right], 
\label{sch_tensor}
\end{eqnarray}
where $4T^3\equiv 4g^{-2}\Lambda^3$ is the 
tension of the wall and $k=\kappa^2T^3$. 
Although our model contains three parameters 
$\Lambda,\ g$ and $\kappa$, this potential 
depends on only two parameters $k$ and $\Lambda$.
If we take the thin wall limit where 
$\Lambda\rightarrow\infty$ fixing $4T^3$, 
we obtain (putting $y_0=0$) 
\begin{eqnarray}
\frac{\Lambda}{\cosh^2\Lambda y(z)} 
\rightarrow 2\delta(z),\quad 
\tanh^2\Lambda y(z) \rightarrow 1,\quad
\left[\cosh\Lambda y(z)
\right]^{-\frac{k}{\Lambda}} 
\rightarrow
 \frac{1}{\left(k|z|+1\right)^2},
\end{eqnarray}
with $kz={\rm sgn}(y){\rm e}^{k|y|}-1$. 
Thus the 
Schr\"odinger potential (\ref{sch_tensor}) 
becomes precisely the potential of the 
Randall-Sundrum model :
\begin{eqnarray}
\mathcal{V}_t(z) \rightarrow 
\frac{2k^2}{\left(k|z| + 1\right)^2}
- 2k\delta(z).
\end{eqnarray}

We find that the part of action quadratic in 
$\tilde{h}^{\rm TT}_{\mu\nu}$ has no $z$ dependent 
weight 
\begin{eqnarray}
S \sim \int dzd^{D-1}x\ \eta^{\mu\rho}
\eta^{\nu\lambda}\tilde{h}^{\rm TT}_{\mu\nu}
\left(\square_{D-1}+\partial_z^2
-\frac{1}{2}\mathcal{V}_t\right)
\tilde{h}^{\rm TT}_{\rho\lambda},
\label{eq:quadratic-h-action}
\end{eqnarray}
in conformity with the absence of the linear term 
\cite{CEHS} 
in $\partial_z$ 
in the Shr\"odinger type equation (\ref{eq:schrod-tt}). 
We stress that this is written in terms of 
the conformal coordinate $z$ and the 
redefined field $\tilde{h}^{\rm TT}_{\mu\nu}$. 

Defining mode equations by eigenvalue equations 
$\mathcal{H}_t
\psi_n(z)=
\left[-\partial_z^2 + \mathcal{V}_t(z)\right]
\psi_n(z)=m_n^2\psi_n(z)$  with 
mass squared eigenvalues $m_n^2$, 
and assuming mode functions $\psi_{n}(z)$ to form 
 a complete set, 
the transverse traceless fields can be expanded into 
a set of effective fields 
$\hat{h}^{{\rm TT} (n)}_{\mu\nu}(x)$ 
\begin{eqnarray}
\tilde{h}^{\rm TT}_{\mu\nu}(x,z) 
= \sum_n \hat{h}^{{\rm TT} (n)}_{\mu\nu}(x)\psi_{n}(z). 
\end{eqnarray}
Then the above quadratic action (\ref{eq:quadratic-h-action}) 
becomes 
\begin{eqnarray}
S \sim \sum_{n, k}\int dz\ \psi_n(z) \psi_k(z) \ \cdot
\int d^{D-1}x\ \eta^{\mu\rho}
\eta^{\nu\lambda}
\hat{h}^{\rm TT(n)}_{\mu\nu}
\left[\square_{D-1}-m_k^2\right]
\hat{h}^{\rm TT(k)}_{\rho\lambda} . 
\end{eqnarray}
Therefore the inner product for the mode function $\psi(z)$ 
should be defined as 
as 
\begin{eqnarray}
\langle\psi_1|\psi_2\rangle = \int dz\ \psi_1(z)\psi_2(z),
\label{eq:inner-prod}
\end{eqnarray}
for which the usual intuition of 
quantum mechanics works.

The Hamiltonian $\mathcal{H}_t$ can now be expressed 
in a SQM form as follows
\begin{eqnarray}
\mathcal{H}_t = Q_t^\dagger Q_t,
\qquad 
Q_t\equiv - \partial_z + {D-2 \over 2}A', 
\qquad 
Q_t^\dagger
\equiv \partial_z + {D-2 \over 2}A', 
\label{eq:SQM-TT}
\end{eqnarray}
where the ``supercharge'' $Q_t$ and $Q_t^\dagger$ 
are adjoint of each other at least for BPS background 
where no boundary condition has to be imposed. 
Therefore the Hamiltonian
$\mathcal{H}_t$ is a nonnegative definite Hermitian 
operator\footnote{
Adjoint relation between $Q_t$ and $Q_t^\dagger$ and 
the Hermiticity of
$\mathcal{H}_t$ are assured by the inner product defined 
in Eq.(\ref{eq:inner-prod}) without $z$ dependent weight. 
}, and its 
eigenvalues are nonnegative definite. 
Therefore we can 
conclude that the  tensor perturbation has 
no tachyonic modes which destabilize the background 
field configurations at least for 
BPS 
solutions. 

There are two possible zero modes 
in the tensor perturbation. 
One is the state which is annihilated by 
$Q_t|\tilde{h}^{\rm TT(+)}_{\mu\nu}\rangle=0$, 
and another is the state defined as 
$Q_t^\dagger\left(Q_t|\tilde{h}^{\rm TT(-)}_{\mu\nu}
\rangle\right)=0$ 
where $Q_t|\tilde{h}^{\rm TT(-)}_{\mu\nu}\rangle\neq0$. 
Then zero modes 
are of the form : 
\begin{alignat}{2}
&\tilde{h}^{\rm TT(0)}_{\mu\nu}(x,z) 
= \hat{h}^{\rm TT(+)}_{\mu\nu} (x)\ 
{\rm e}^{{D-2 \over 2}A(z)}
+ \hat{h}^{\rm TT(-)}_{\mu\nu} (x)\ 
{\rm e}^{{D-2 \over 2}A(z)}\int dz\ {\rm
 e}^{-(D-2)A(z)},
 \label{eq:tensor-zero-mode}
\end{alignat}
where $A(z) = A(y(z))$. 
Notice that 
we must verify the normalizability of the wave-function 
to obtain a physical massless effective field in the case of 
noncompact space such as our BPS background. 
In the case of non-BPS background, the 
boundary condition has to be verified, which we shall 
consider in Sec.\ref{sc:stability=nonBPS}. 
The first mode $\hat{h}^{\rm TT(+)}_{\mu\nu}$ 
in Eq.(\ref{eq:tensor-zero-mode}) is 
normalizable 
if $\int dz\ {\rm e}^{(D-2)A(z)} < \infty$, 
corresponds to the graviton which is 
localized at the wall with a positive energy density. 
namely, if 
${\rm e}^{(D-2)A}$ falls off faster than $|z|^{-1}$ 
\cite{CEHS}. 
 For our BPS solution
(\ref{BPS_sol}) the asymptotic behavior of the warp 
factor ${\rm e}^A$ is of order
$|z|^{-1}$. 
Therefore we obtain a normalizable 
massless transverse traceless mode 
$\hat{h}^{\rm TT(+)}_{\mu\nu}$ which gives 
the physical graviton localized on the wall. 

On the other hand, the second term 
 $\hat{h}^{\rm TT(-)}_{\mu\nu}$ 
 in Eq.(\ref{eq:tensor-zero-mode}) 
is not normalizable and is 
unphysical since $\left({\rm e}^{D-2\over 2}A\int dz\
{\rm e}^{-(D-2)A}\right)^2\sim |z|^4$ at 
$|z| \rightarrow \infty$ for our BPS solution.
If there exists a regulator brane with a negative 
tension at some $y$, this mode can become normalizable
and localizes at the negative tension brane 
in contrast to the graviton. 
If there are no bulk scalar fields 
(contrary to our model) as in the original 
Randall-Sundrum model of single wall, 
this zero mode corresponds 
to the physical massless field which was called 
radion in Ref.\cite{Charmousis:1999rg}. 

Our specific four-dimensional model of non-BPS wall 
gives a three-dimensional effective theory on the wall. 
Transverse traceless mode of graviton in three dimensions 
has no dynamical degree of freedom except possible 
topological modes. 
However, our formalism and analysis can be applied 
at each step to general $D$-dimensional theories, 
once we obtain 
the relevant non-BPS solutions in such theories. 
In that respect, we believe that our findings should 
still be useful. 

%%%%%%%%%%%%%%%%%%%%%%%%%%%%%%%%%%%%%%%%%%%%%%%%%%%%%%%%%%%%%%%%%%
\begin{figure}[htb]
\begin{center}
\includegraphics[width=6.5cm]{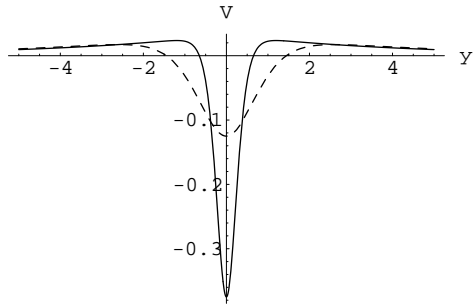}
 \caption{{\sf {\small The Schr\"odinger potential for the tensor
 perturbation in terms of the coordinate system $y$. We choose $y_0=0$
 and $k=0.25$. The solid line has $\Lambda =3$ and
 the broken line has $\Lambda=1$.}}}
\label{V_tensor}
\end{center}
\end{figure}
%%%%%%%%%%%%%%%%%%%%%%%%%%%%%%%%%%%%%%%%%%%%%%%%%%%%%%%%%%%%%%%%%%

The Sch\"odinger potential
can always be 
expressed in terms of $y$, 
but is difficult in terms of $z$ explicitly\footnote{
For a special case where $k\Lambda^{-1}$ is an integer, we can express
the Schr\"odinger potential in terms of $z$ explicitly. We show this in
Appendix B}, 
since 
it is generally difficult to solve 
$dz = {\rm e}^{-A}dy$ explicitly. 
If we express the potential in terms of $y$, 
we obtain a volcano type potential as shown 
in Fig.\ref{V_tensor}. 
The width of the well is $\sim 2\Lambda^{-1}$ 
and the depth is $\sim k\Lambda$.

Next we turn to analysis of the massive Kaluza-Klein 
(KK) mode. 
There are no modes with 
negative mass squared in the tensor
perturbation, 
as we have already shown. 
Since the Schr\"odinger potential 
(\ref{sch_tensor}) 
vanishes asymptotically ($z=\pm\infty$), 
all the massive KK modes are continuum scattering states 
with eigenvalues $m^2>0$.  
In order to examine the mode functions of the massive 
KK modes, we look into the region 
far from the wall, namely $\Lambda|y|\gg1$. 
Since ${\rm e}^{A} \simeq {\rm e}^{-k|y|}$, 
$kz\simeq {\rm sgn}(y){\rm e}^{k|y|}-1$, 
we find that the Schr\"odinger potential becomes 
\begin{eqnarray}
\mathcal{V}_t(z) \simeq 
\frac{2k^2}{\left(k|z| + 1\right)^2}
\qquad \left(\Lambda|y|\gg1\right). 
\label{large_y_tensor}
\end{eqnarray}
This happens to be the same potential as that in the 
Randall-Sundrum single wall model \cite{RS2}, 
in spite of different spacetime dimensions. 
The wave functions of the continuum massive modes 
for this potential are known to be expressed as 
linear combinations of Bessel functions at the 
region far from the wall \cite{RS2}.

\subsection{
The active scalar perturbation}
\label{sc:active-scalar}

Next we study the perturbation of the active scalar 
field $\varphi_{\rm R}$. 
Notice that the fluctuation $\varphi_{\rm R}$ 
around the active scalar field background $\phi_{\rm R}$ 
can be reduced to the trace part $B$ of the 
metric perturbation through Eq.(\ref{Newton_3}). 
Therefore we mainly concentrate on 
the trace (scalar) part of the metric 
perturbation $B$ in what follows. 
The linearized equation 
which contains only $B$ can be derived by combining 
Eq.(\ref{Newton_2}) and 
(\ref{Newton_3}) and using the background field equation : 
\begin{eqnarray}
\left[{\rm e}^{-2A}\square_{D-1} + \partial_y^2
+ \left((D-3)\dot{A}-2\frac{\ddot\phi_{\rm R}}
{\dot\phi_{\rm R}}
\right)\partial_y
+ 2(D-3)\left(\ddot{A}-\dot{A}
\frac{\ddot\phi_{\rm R}}{\dot\phi_{\rm R}}\right)
\right]B=0.\label{B_eq}
\end{eqnarray}
In order to transform this into the Schr\"odinger 
form, we change the
coordinate from $y$ to $z$ and redefine the field as
$\tilde{B}\equiv{\rm e}^{{D-2 \over 2}A}
\phi_{\rm R}'{^{-1}}B$. 
Substituting this into
Eq.(\ref{B_eq}), we find the Sch\"odinger type equation 
for the scalar perturbation;
\begin{eqnarray}
\mathcal{H}_e \tilde{B} \equiv \left[-\partial_z^2 
+ \mathcal{V}_e(z)\right]\tilde{B} 
= \square_{D-1}\tilde{B}, 
\label{Ham_scalar}
\end{eqnarray}
where the Schr\"odinger potential $\mathcal{V}_e(z)$ 
is defined by 
\begin{eqnarray}
\mathcal{V}_e(z) \equiv 
- \frac{\phi_{\rm R}'''}{\phi_{\rm R}'} 
+ 2 \left(\frac{\phi_{\rm R}''}{\phi_{\rm R}'}\right)^2
+(D-4){A''\phi_R'' \over \phi_R'}
- {3D-10 \over 2}A'' + \left({D-2 \over 2}\right)^2A'{^2}.
\label{eq:scalar-fluc-pot}
\end{eqnarray}
Similarly to the tensor perturbation, 
the inner-product for the scalar 
perturbation $B$ should be defined in terms of the 
conformal coordinate $z$ 
and the redefined field $\tilde{B}$. 

Plugging our solution (\ref{BPS_sol}) into this, we find 
\begin{eqnarray}
\mathcal{V}_e = 
\left[\cosh\Lambda (y-y_0)\right]^{-\frac{2k}{\Lambda}}
\left[
\Lambda^2 + k\Lambda
\left(1+\frac{1}{\cosh^2\Lambda (y-y_0)}\right)
\right]. 
\label{Sch_scalar}
\end{eqnarray}
%%%%%%%%%%%%%%%%%%%%%%%%%%%%%%%%%%%%%%%%%%%%%%%%%%%%%%%%%%%
\begin{figure}[t]
\begin{center}
\includegraphics[width=6.5cm]{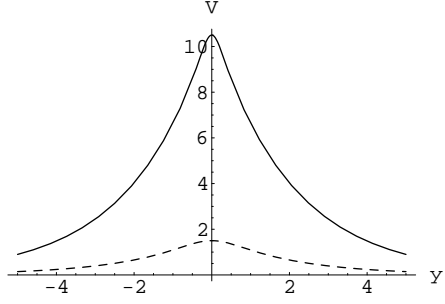}
 \caption{{\sf {\small The Schr\"odinger potential for 
 the scalar 
 perturbation in terms of the coordinate system $y$. 
 We choose $y_0=0$, 
 and $k=0.25$. The solid line has $\Lambda =3$ and 
 the broken line has $\Lambda=1$.}}}
\label{V_scalar}
\end{center}
\end{figure}
%%%%%%%%%%%%%%%%%%%%%%%%%%%%%%%%%%%%%%%%%%%%%%%%%%%%%%%%%%
We stress that $\mathcal{V}_e$ can be expressed 
in terms of $y$, but not in terms of $z$, 
since it is generally difficult to solve 
$dz={\rm e}^{-A}dy$. 
This potential $\mathcal{V}_e$ has the following 
properties : 
 i) it is positive definite, 
 ii) it vanishes asymptotically 
 at infinity, and 
 iii) the height of $\mathcal{V}_e$ is of order 
$\Lambda^2$ as shown in Fig.\ref{V_scalar}. 
From i) , it follows that there are no tachyonic modes 
since the wave function of such 
modes  will necessarily diverge either at 
$y=\infty$ or $y=-\infty$. 
Therefore we can conclude that the background configuration 
(\ref{BPS_sol}) 
is stable under the active scalar perturbation. 
From ii), it follows that 
the spectrum of the massive modes is 
continuous starting from zero. 
From iii), the potential diverges at any finite 
point $y$ in the thin wall limit 
$(\Lambda\rightarrow\infty)$.

Though we can not find the exact solutions for 
the massive KK modes,
zero modes can be found by rewriting 
the Hamiltonian (\ref{Ham_scalar})
into SQM form as follows : 
\begin{eqnarray}
\mathcal{H}_e = Q_e^\dagger Q_e,
\quad\quad 
Q_e \equiv 
-\partial_z + 
\left[\log\left({\rm e}^{-{D-2 \over 2}A}
\dfrac{A'}{\phi_{\rm R}'}\right)\right]',
\quad
Q_e^\dagger \equiv 
\partial_z + 
\left[\log\left({\rm e}^{-{D-2 \over 2}A}
\dfrac{A'}{\phi_{\rm R}'}\right)\right]'.
\label{eq:SQM-scalar}
\end{eqnarray}
To show this, we use the identity 
$\phi_{\rm R}'\left(A'''-2A'A''\right) =
2\phi_{\rm R}''\left(A''-A'{^2}\right)$.
Similarly to the tensor perturbation there are 
two zero modes 
of $\mathcal{H}_e$ :  
\begin{alignat}{2}
&\tilde{B}^{(0)}(x,z) 
= \hat{B}^{(+)}(x)\ \frac{A'}{\phi_{\rm R}'}
{\rm e}^{-{D-2 \over 2}A}
+ \hat{B}^{(-)}(x)\ \frac{A'}{\phi_{\rm R}'}
{\rm e}^{-{D-2 \over 2}A}
\int dz\ \frac{\phi_{\rm R}'{^2}}{A'{^2}}
{\rm e}^{(D-2)A} .
\label{zero_scalar}
\end{alignat}
Both zero modes are unphysical by the following reasons. 
The first term is unphysical, 
in the sense that this is 
eliminated by a gauge transformation preserving 
the Newton gauge (\ref{eq:newton-gauge}) : 
\begin{eqnarray}
\xi_2 = \hat{B}^{(+)}(x){{\rm e}^{-(D-3)A} \over D-3},
\quad 
\xi_\mu 
= - \hat{B}^{(+)}_{,\mu}(x)
{{\rm e}^{2A} \over D-3}\int dy\ {\rm e}^{-(D-1)A},
\label{gauge_transf_newton}
\end{eqnarray}
where $\xi_m$ is an infinitesimal coordinate transformation
parameters. The transformation law is given in Appendix A.
The second term is unphysical 
since it diverges at infinity and is not normalizable 
as illustrated in Fig.\ref{V_scalar_z}.

Next we turn to analysis for the massive KK modes. 
As we have mentioned
above, massive modes are continuous from zero. 
Similarly to the tensor perturbation, properties of 
mode functions can be examined by analyzing the 
behavior of the potential in the region far from the wall. 
In the region where $|y|\Lambda\gg1$, the Schr\"odinger 
potential (\ref{Sch_scalar}) becomes 
 : 
\begin{eqnarray}
\mathcal{V}_e(z) \simeq 
\frac{\Lambda^2 + k\Lambda}{\left(k|z| +1\right)^2}. 
\label{large_y_scalar}
\end{eqnarray}
This potential is very similar to the Schr\"odinger 
potential (\ref{large_y_tensor}) 
for the tensor perturbation. 
Therefore all the massive modes are 
given by a linear combination 
of Bessel functions asymptotically at 
$|z|\rightarrow\infty$.
Although these two Schr\"odinger potentials 
(\ref{large_y_tensor}) and
(\ref{large_y_scalar}) have the same $z$ dependence 
asymptotically $|y|\Lambda\gg1$,
their behaviors in the thin wall limit are very different. 
The potential (\ref{large_y_tensor}) depends only on 
$k$ (fixed in the thin-wall limit), but not on $\Lambda$. 
On the other hand, the potential (\ref{large_y_scalar}) 
is proportional to polynomials in $\Lambda$. 
Therefore, the latter diverges in thin wall limit whereas 
the former is finite. 
This can be understood 
as follows. 
The perturbation of the trace part of the metric $B$ 
is related to the active scalar 
field perturbation $\varphi_{\rm R}$ 
through Eq.(\ref{Newton_3}). 
Since all the massive KK modes associated with 
the active scalar field become infinitely heavy 
in the thin wall limit, the massive KK 
modes for the perturbation of the trace part 
of the metric freeze simultaneously. 
In this limit 
only the tensor perturbations remain which 
correspond 
to the known modes of the RS model\footnote{
Generically speaking, the fluctuations 
of the inert scalar field $\varphi_{\rm I}$ 
can be an exception depending on the potential, 
although the inert scalar $\varphi_{\rm I}$ 
in our model is also frozen in the thin wall limit. 
}
.

The zero modes 
$\hat{B}^{(+)}(x)\dfrac{A'}{\phi_{\rm R}'}
{\rm e}^{-{D-2 \over 2}A}$ 
of the fluctuation of the trace part of 
the metric $B$ in Eq.(\ref{zero_scalar}) 
can be translated into 
the perturbation of the active scalar field 
$\varphi_{\rm R}$ by means of Eq.(\ref{Newton_3}) 
: 
\begin{eqnarray}
\varphi_{\rm R}^{(0)}(x,y) 
= \hat\varphi_{\rm R}^{(+)}(x)\ \dot\phi_{\rm R}
{\rm e}^{-(D-3)A}
\rightarrow 
\hat\varphi_{\rm R}^{(+)}(x)\ \dot\phi_{\rm R},\quad
\left(\kappa\rightarrow0\right).
\end{eqnarray}
where $\hat\varphi_{\rm R}^{(+)}(x)\equiv 
(D-2)\hat{B}_{\rm R}^{(+)}(x)/2$. 
In weak gravity limit $(\kappa\rightarrow0)$, 
${\rm e}^{A}$ reduces to
a constant. 
Then we find that this zero mode is localized on the wall 
and that it corresponds 
to the Nambu-Goldstone boson 
corresponding to the spontaneously broken translational 
invariance.

\subsection{
Analysis for the perturbation 
about $\phi_{\rm I}$}

In our tree level solution the imaginary part 
of the scalar field $\phi$ vanishes identically 
and does not contribute to the energy
momentum tensor. 
Therefore it does not 
affect the spacetime
geometry. 
We shall call scalar fields with no nontrivial field 
configuration as inert field. 
In the linear order of perturbations we found that 
the fluctuation $\varphi_{\rm I}$ of this inert field 
 decouples from any other fluctuations, as shown in 
Eq.(\ref{inert}).

In order to find the spectrum of 
$\varphi_{\rm I}$, 
 we first bring  Eq.(\ref{inert}) into a 
Schr\"odinger form by changing the coordinate from $y$ 
to $z$ and redefining the field 
$\tilde\varphi_{\rm I}\equiv
{\rm e}^{{D-2 \over 2}A}\varphi_{\rm I}$.
Then we obtain 
\begin{eqnarray}
\mathcal{H}_{\rm I}\tilde\varphi_I
\equiv \left[-\partial_z^2 
+ \mathcal{V}_{\rm I}(z)\right]\tilde\varphi_{\rm I} 
= \square_{D-1}\tilde\varphi_{\rm I},
\quad
\mathcal{V_{\rm I}}(z)
\equiv \mathcal{V}_t(z) 
+ {\rm e}^{2A}\dfrac{1}{2}
\dfrac{\partial^2V}{\partial\phi_{\rm I}^2}\bigg|.
\label{eq:inert-hamilton}
\end{eqnarray}
where ${\cal V}_t(z)$ is the potential for transverse 
traceless part of the metric defined in 
Eq.(\ref{eq:schrod-tt}). 
To obtain more concrete informations on the spectrum, 
we need to examine properties of each model. 
For our model we find 
\begin{eqnarray}
\frac{1}{2}\frac{\partial^2V}{\partial\phi_{\rm I}^2}\bigg| 
&=& \Lambda^2
+ \frac{\kappa^2\Lambda^4}{g^2}
\left(1+2\cos^2\frac{g}{\Lambda}\phi_{\rm R}\right)
- \frac{2\kappa^4\Lambda^6}{g^4}
\sin^2\frac{g}{\Lambda}\phi_{\rm R}
\label{eq:inert-potential}
\end{eqnarray}
We shall discuss generic property of this inert scalar 
for the non-BPS background in 
Sec.\ref{sc:stability=nonBPS}. 

\begin{figure}[htb]
\begin{center}
\includegraphics[width=6.5cm]{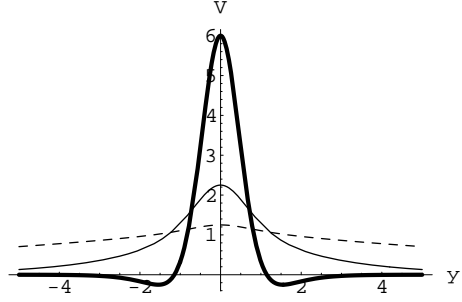}
 \caption{{\sf {\small The Schr\"odinger potential for the perturbation
 of the imaginary part of the scalar field in terms of the coordinate
 system $y$. We choose $y_0=0$
 and $\Lambda=1$. The bold line has $k=1$, the solid line has $k=0.25$ and
 the broken line has $k=0.05$.}}}
\label{V_inert}
\end{center}
\end{figure}

If we choose the BPS solution as our background, 
we can rewrite the potential by using the BPS 
equations (\ref{BPS_eq_R}) 
\begin{eqnarray}
\frac{1}{2}\frac{\partial^2V}{\partial\phi_{\rm I}^2}\bigg| 
&=& \Lambda^2 + 2k\Lambda - 2\ddot{A} - 2\dot{A}^2,
\end{eqnarray}
Then the Schr\"odinger potential $\mathcal{V}_{\rm I}$ 
takes the form : 
\begin{eqnarray}
\mathcal{V}_{\rm I} = A'{^2} - A'' 
+ {\rm e}^{2A}\left(\Lambda^2 + 2k\Lambda\right).
\end{eqnarray}
We illustrate $\mathcal{V}_I$ in terms of $y$ 
in Fig.\ref{V_inert}.
 For vanishing gravitational coupling $\kappa \rightarrow 0$, 
 $\mathcal{V}_{\rm I}$ reduces to a constant 
$\Lambda^2$, which agrees with the model of  global 
SUSY in Ref.\cite{EMSS}. 
On the other hand, the potential $\mathcal{V}_{\rm I}$ 
 acquires regions of negative values when 
$\kappa$ becomes large. 
%%%%%%%%%%%%%%%%%%%%%%%%%%%%%%%%%%%%%%%%%%%%%%%%%%%%%%%%%%%%%%%%%%
%%%%%%%%%%%%%%%%%%%%%%%%%%%%%%%%%%%%%%%%%%%%%%%%%%%%%%%%%%%%%%%%%%
In the case of the BPS background, 
we can show that there are no tachyonic modes 
in this inert scalar sector with 
the aid of the SQM. 
Let us introduce a supercharge as 
$Q_{\rm I}
= -\partial_z - A'$ and $Q_{\rm I}^\dagger=\partial_z - A'$. 
Then the
Hamiltonian $\mathcal{H}_{\rm I}$ can be rewritten as
\begin{eqnarray}
\mathcal{H}_{\rm I} = Q_{\rm I}^\dagger Q_{\rm I} 
+ {\rm e}^{2A}\left(\Lambda^2 + 2k\Lambda\right).
\end{eqnarray}
The first term is a nonnegative definite Hermitian operator 
and the second 
term is never negative. 
Therefore, we can conclude that
eigenvalues of $\mathcal{H}_{\rm I}$ are always nonnegative 
and there are no tachyonic mode.

\section{Stability of Non-BPS multi-Walls }
\label{sc:stability=nonBPS}

For non-BPS solutions, the positivity of the energy 
of the fluctuation and the associated stability is 
entirely nontrivial. 
In the limit of vanishing gravitational coupling, 
however, our supergravity model reduces to a global 
SUSY model that has been shown to be stable \cite{EMSS}. 
Since the mass gap in the global SUSY model should 
not disappear even if we switch on the gravitational 
coupling infinitesimally, the massive scalar fluctuations 
in the global SUSY model should remain massive at least 
for small gravitational coupling. 
On the other hand, we need to watch out a possible new 
tachyonic instability associated with the metric 
fluctuations. 

As for the transverse traceless part of the metric, 
we have already shown that there are two possible 
zero mode candidates 
$ \hat{h}^{\rm TT(+)}_{\mu\nu} (x)\ {\rm e}^{A(z)}$ 
and 
$\hat{h}^{\rm TT(-)}_{\mu\nu} (x)\ 
{\rm e}^{A(z)}\int dz\ {\rm e}^{-2A(z)}$ 
in Eq.(\ref{eq:tensor-zero-mode}). 
In the non-BPS solution, we no longer need to worry about 
the normalizability of the wave function. 
Instead, we need to satisfy the boundary condition 
imposed by the presence of the boundary cosmological 
constants. 
To impose the boundary condition, we have to use coordinate 
system which is more more appropriate to specify the 
position of the boundary. 
This is achieved by going to the Gaussian normal 
coordinates \cite{TanakaMontes}. 
We have been using the Newton gauge to study the mass 
spectrum in the Shr\"odinger type equation. 
We can follow the argument in Ref.\cite{TanakaMontes} 
to obtain general coordinate transformations 
$\xi_m$ from the Newton 
gauge to the Gaussian normal gauge :  
\begin{equation}
\xi_2(x,y)={1 \over 2}\int_{0}^{y} dy' h_{22} 
+ \bar \xi_2^{(\pm)}(x), 
\end{equation}
\begin{equation}
\xi_\mu(x,y)=
-{1 \over 2}\int_{y^{(\pm)}}^ydy'e^{-2A}
\int_{y^{(\pm)}}^{y'}dy'' h_{22, \mu}
-\bar \xi_{2,\mu}^{(\pm)}\int_{y^{(\pm)}}^ydy'e^{-2A}
+ \bar \xi_\mu^{(\pm)}(x), 
\end{equation}
where $\bar \xi_2^{(\pm)}, \bar \xi_\mu^{(\pm)}$ 
depend on $x$ only. 
Boundary conditions in the Newton gauge are found to be \cite{TanakaMontes}
\begin{equation}
\left[\partial_y h_{\mu\nu}^{TT}
+2e^{-2A}\bar \xi_{2,\mu,\nu}^{(\pm)}
\right]_{y^{(\pm)}-0}^{y^{(\pm)}+0}
=0
, \label{const_tensor}
\end{equation}
\begin{equation}
\left[\left(\partial_y+\dot A\right) h_{22} 
+ 2\ddot{A} \bar \xi_2^{(\pm)}
\right]_{y^{(\pm)}-0}^{y^{(\pm)}+0}
=0. \label{const_scalar}
\end{equation}
We find that the former mode 
$ \hat{h}^{\rm TT(+)}_{\mu\nu} (x)\ {\rm e}^{A(z)}$ 
receives no constraint from the boundary condition, 
and is still a physical massless mode 
localized on the wall which should be regarded as 
the graviton in the effective theory. 
The latter mode 
$\hat{h}^{\rm TT(-)}_{\mu\nu} (x)\ 
{\rm e}^{A(z)}\int dz\ {\rm e}^{-2A(z)}$ 
is constrained by the boundary condition 
(\ref{const_tensor}).
Since the constraint (\ref{const_tensor}) and 
(\ref{const_scalar}) relate
$\bar \xi_2^{(\pm)}$ to $\hat h^{(-)}_{\mu\nu}$ and
$\hat h_{22}(x)$, 
this mode should be classified as a scalar type perturbation.
On the other hand, $\hat h^{(-)}_{\mu\nu}$ can be gauged 
away through the gauge transformation 
(\ref{gauge_transf_newton}), which preserves 
the Newton gauge\footnote{
For simplicity, we have assumed $Z_2$ parity of the 
metric perturbation to be even. 
}.
Therefore this mode becomes unphysical in the 
presence of the bulk scalar field like in our model. 
Since $\bar \xi_2$ corresponds to the physical
distance between the two branes 
\cite{Charmousis:1999rg}, our system automatically 
incorporates the stabilization mechanism without an 
additional bulk scalar fields.
In the thin wall limit, the wall scalar field freezes 
out and the fluctuation of the scalar field 
ceases to be related to the scalar perturbation 
of the metric, so that $\hat h_{\mu\nu}^{(-)}$ can no 
longer be gauged away. 
As a result of restoration of $\hat h_{\mu\nu}^{(-)}$, any 
distance between two walls is admitted as classical solutions 
and the model becomes meta-stable.

To evaluate the mass spectrum of massive modes, 
we use the small width approximation. 
Then the asymptotic behavior of the potential 
gives the wave functions expressed by means of 
the Bessel functions as in the Randall-Sundrum 
model. 
These eigenvalue spectrum is approximately 
equally spaced just like the plane wave solutions. 
These (almost) continuum modes should give the corrections 
to the effects of localized graviton, similarly to the 
Randall-Sundrum model. 
Therefore we obtain a massless graviton localized 
on the wall and a tower of massive KK modes 
for transverse traceless part of the metric.

Since the active scalar field $\varphi_{\rm R}$ 
exhibits a mass gap without any tachyon, 
we expect that there should be no tachyonic instability 
at least for small enough gravitational coupling. 
We have found in Eq.(\ref{Newton_3}) that 
the active scalar field $\varphi_{\rm R}$ can be reduced 
to the trace part of the metric $B$. 
This implies that there should be no tachyon 
in the transverse traceless mode as well at least 
for small gravitational coupling, since 
both degrees of freedom represent one and the same 
dynamical degree of freedom. 
In fact, we have observed that the potential 
${\cal V}_{e}$ defined in Eq.(\ref{eq:scalar-fluc-pot}) 
is everywhere positive and has no tachyon. 

There are two possible zero mode candidates 
as given in Eq.(\ref{zero_scalar}). 
However, both of them are unphysical by the following 
reason. 
The first one can be eliminated by a gauge transformation. 
The second one has now  no problem of normalization, since 
the extra dimension is now a finite interval, but 
it cannot satisfy the boundary conditions \cite{TanakaMontes}. 

To evaluate the mass of the lightest scalar particle, which is 
usually called radion, we use the thin-wall approximation where 
the wall width is assumed to be small compared to the radius of 
compactification $R$. 
To make this approximation, we separate the potential 
(\ref{eq:scalar-fluc-pot}) for the trace part of the metric 
fluctuation 
into unperturbed and perturbation as 
\begin{equation}
\mathcal{V}_e(z) \equiv 
\mathcal{V}_e^{(0)}(z) + 
\mathcal{V}_e^{(1)}(z) ,
\end{equation}
\begin{equation}
\mathcal{V}_e^{(0)}(z) = 
\phi_{\rm R}' e^{{D-4 \over 2}A}\left({ e^{-{D-4 \over 2}A} 
\over \phi_{\rm R}'}\right)'', 
\qquad 
\mathcal{V}_e^{(1)}(z) = 
{D-3 \over D-2} 2 \kappa^2 \left(\phi_{\rm R}'\right)^2, 
\end{equation}
where we considered in $D$-dimensions instead of 
$4$ dimensions. 
The zero-th order eigenfunction for the lowest eigenvalue 
is found to be 
\begin{equation}
\tilde B^{(0)}(z)=
{{\cal N} \over {\rm e}^{{D-2 \over 2}A}\phi_{\rm R}'}
,
\end{equation}
with the vanishing eigenvalue $(m_0^{(0)})^2=0$ 
and a normalization factor ${\cal N}$. 
The first order eigenfunction is given by 
\begin{equation}
\tilde B^{(1)}(z)=
{{\cal N} \over {\rm e}^{{D-4 \over 2}A}\phi_{\rm R}^{'}}
\int^z dz' {\rm e}^{(D-4)A} \phi_{\rm R}^{'2}
\int^{z'} dz'' {\rm e}^{-(D-4)A} \phi_{\rm R}^{'-2}
\left(\mathcal{V}_e^{(1)}(z)
-(m_0^{(1)})^2\right)
,
\end{equation}
where the first correction to the mass squared eigenvalue 
is denoted as $(m_0^{(1)})^2$. 
By using Eq.(\ref{Newton_3}), the trace part of the metric can be 
transformed into the active scalar fluctuation $\varphi_{\rm R}$ 
as 
\begin{equation}
\varphi_0^{(1)}(y)=-{D-2 \over \kappa^2}
{\cal N}  \dot\phi_{0} 
\int_{0}^{y_{(+)}} dy' 
\left({D-3 \over D-2}\kappa^2 {\rm e}^{-(D-3)A} 
-{(m_0^{(1)})^2{\rm e}^{-(D-1)A} \over 2 \dot\phi_{\rm R}^{2}}
\right)
,
\end{equation}
where the position of the wall is at $y=0$ and 
the boundary with the negative 
cosmological constant is $y_{(+)}$. 
To satisfy the correct 
boundary conditions, we have to require that 
this first order eigenfunction $\varphi_0^{(1)}$ 
should vanish at the boundary 
\cite{TanakaMontes}. 
This determines the first order mass squared as 
\begin{equation}
(m_0^{(1)})^2=2\kappa^2 
\frac{D-3}{D-2}
\frac{\displaystyle \int_0^{y_{(+)}} dy\ {\rm e}^{-A}}{
\displaystyle \int_{0}^{y_{(+)}} dy\ 
\dfrac{{\rm e}^{-3A}}{\dot \phi_{0}^2}},
\end{equation}

Taking $D=4$ as in our model, and applying to 
the case of the first example of non-BPS background 
in Eq.(\ref{eq:negaive-cosm-pi/2}) with the symmetry 
$Z_2\times Z_2$, 
we should identify $y_{(+)}=\pi R/2$ and obtain 
\begin{equation}
m^2_0 \approx 8\Lambda^2 
e^{-(1+\alpha^2)\pi \Lambda R} 
\left(1+{3 \over 2}\alpha^2\right)
2^{\alpha^2}, 
\qquad 
\alpha^2 \equiv 
{\kappa^2 T^3 \over \Lambda },
\label{eq:radion-mass}
\end{equation}
where $4T^3$ is the tension (energy density) of the wall. 
For the other background solution with the $Z_2\times Z_2$ 
symmetry, we should identify the boundary with the negative 
cosmological constant as $y_{(+)}=\pi R$, and obtain 
the same result. 
It is interesting to note 
that the mass scale is given by the inverse 
wall width $\Lambda$, and that it 
becomes exponentially light as a 
function of the distance between the walls, 
even though the radion mass receives a complicated 
gravitational corrections. 
It is appropriate to fix 
 the wall tension 
and the gravitational coupling 
in taking the small width limit 
$\Lambda R \rightarrow \infty$. 
Then we obtain a simple mass formula in the 
limit \footnote{
This mass squared is factor two larger compared to the 
value of lightest massive scalar in the global SUSY 
model\cite{MSSS2}. 
We have not understood this discrepancy. 
}
\begin{equation}
m^2_0 \approx 8\Lambda^2 
e^{-\pi \Lambda R} \rightarrow 0 . 
\label{eq:radion-mass2}
\end{equation}
This characteristic feature of lightest massive scalar 
fluctuation is precisely the same as the global SUSY 
case \cite{MSSS2}. 
The lightest massive mode in that case results from the fact 
that two walls have no communication when they are far apart, 
and the translation zero modes of each wall becomes 
massless as the separation between walls goes to infinity.

The mass spectrum of the inert scalar fluctuations 
$\varphi_{\rm I}$ is determined by the Schr\"ordinger 
form of the eigenvalue problem (\ref{eq:inert-hamilton}) 
with the potential $\mathcal{V_{\rm I}}(z)$. 
The potential has the same term as the transverse 
traceless mode $\mathcal{V_{\rm t}}(z)$ 
with an additional term 
$\frac{1}{2}\frac{\partial^2V}{\partial\phi_{\rm I}^2}\bigg|$ 
in Eq.(\ref{eq:inert-potential}) which is nonnegative 
definite provided the gravitational 
coupling is not too strong $\kappa \le {g \over \Lambda}$ 
\begin{eqnarray}
\frac{1}{2}\frac{\partial^2V}{\partial\phi_{\rm I}^2}\bigg| 
&=& \Lambda^2\left(1+3\frac{\kappa^2\Lambda^2}{g^2}\right) 
- 2\frac{\kappa^4\Lambda^4}{g^2}
\left(1+ \frac{\kappa^2\Lambda^2}{g^2}\right)
\sin^2\frac{g}{\Lambda}\phi_{\rm R}
\nonumber \\
&\ge &
 \Lambda^2\left(1-\frac{\kappa^2\Lambda^2}{g^2}\right) 
\left(1+2\frac{\kappa^2\Lambda^2}{g^2}\right) \ge 0
.
\label{eq:inert-positivity}
\end{eqnarray}
Therefore inert scalar does not produce any additional 
tachyonic instability.

%%%%%%%%%%%%%%%%%%
\section{Fermions}
%%%%%%%%%%%%%%%%%%
In the previous two sections, 
 we focused on the stability of BPS and Non-BPS wall 
 configurations 
 and studied its fluctuations. 
In this section, 
 we turn to the fermionic part of the model and study 
 its fluctuation. 
 We shall consider only the BPS solutions for simplicity, 
 since it allows massless gravitino, whereas 
 the non-BPS solutions do not. 
The part of the Lagrangian (\ref{SUGRA_Lag}) 
quadratic in fermion fields (with arbitrary powers 
of bosons) can be rewritten as 
\begin{eqnarray}
\label{fermionlag}
e^{-1}{\cal L}_{{\rm fermion}}^{{\rm quadratic}} &=& 
-i {\rm e}^{-A} \bchi \bsig^{\underline{\mu}} 
{\cal D}_\mu \chi 
-i \bchi \bsig^{\underline{2}} {\cal D}_2 \chi 
+ \varepsilon^{\underline{\kappa 2 \mu \nu}} 
{\rm e}^{-3A} 
\bpsi_{\kappa} \bsig_{\underline{2}} 
\tilde{{\cal D}}_\mu \psi_{\nu} 
+ \vep^{\underline{\kappa \lambda 2 \nu}} 
{\rm e}^{-2A} \bpsi_{\kappa} \bsig_{\underline{\lambda}} 
\tilde{{\cal D}}_2 \psi_{\nu} 
\nonumber \\
&& 
+ \vep^{\underline{2 \lambda \mu \nu}} 
{\rm e}^{-2A} \bpsi_{2} \bsig_{\underline{\lambda}} 
\tilde{{\cal D}}_{\mu} \psi_{\nu} 
+ \vep^{\underline{\kappa \lambda \mu 2}} 
{\rm e}^{-2A} \bpsi_{\kappa} \bsig_{\underline{\lambda}} 
\tilde{{\cal D}}_{\mu} \psi_{2} 
\nonumber \\
&& 
-\frac{\kappa}{\sqrt{2}} {\rm e}^{-A} \dot{\phi}^* \chi 
\sig^{\underline{\mu}} 
\bsig^{\underline{2}} \psi_\mu 
- \frac{\kappa}{\sqrt{2}} {\rm e}^{-A} \dot{\phi} \bchi 
\bsig^{\underline{\mu}} 
\sig^{\underline{2}} \bpsi_\mu 
+ \frac{\kappa}{\sqrt{2}} \dot{\phi}^* \chi \psi_2 
+ \frac{\kappa}{\sqrt{2}} \dot{\phi} \bchi \bpsi_2 
\nonumber \\
&& - \kappa^2 {\rm e}^{{\kappa^2 \over 2}\phi^*\phi}
\biggl[
P^* {\rm e}^{-2A} \psi_\mu \sig^{\underline{\mu \nu}} 
\psi_\nu 
+ P {\rm e}^{-2A} \bpsi_\mu \bsig^{\underline{\mu \nu}} 
\bpsi_\nu 
+ 2 {\rm e}^{-A} 
(P^* \psi_\mu \sig^{\underline{\mu 2}} \psi_2 
+ P \bpsi_\mu \sig^{\underline{\mu 2}} \bpsi_2)  
\nonumber \\
&& +  
 {\rm e}^{-A} \frac{i\kappa}{\sqrt{2}} ( D_\phi P \chi 
\sig^{\underline{\mu}} \bar{\psi}_\mu 
+ D_{\phi^*} P^* \bchi \bsig^{\underline{\mu}} \psi_\mu)
+ \frac{i\kappa}{\sqrt{2}} 
(D_\phi P \chi \sig^{\underline{2}} \bpsi_2 
+ D_{\phi^*} P^* \bchi \bsig^{\underline{2}} \psi_2) 
 \nonumber \\
&&
+  \half ( {\cal D}_\phi D_\phi P \chi^2 
+ {\cal D}_{\phi^*} D_{\phi^*} P^* \bchi^2)
\biggr].
\label{eq:fermion-lag}
\end{eqnarray}
The terms in the fourth line is quadratic in gravitino 
without any derivatives, which can be regarded as 
mass terms for gravitino. 
We find that they are $Z_2$ odd under 
$\phi \rightarrow -\phi$. 
In this respect, our model provides an explicit realization of 
the condition to have a smooth 
limit of vanishing width of the wall \cite{BCY} 
and in agreement with one version of the 
five-dimensional supergravity on the orbifold \cite{FLP}. 
For our modified superpotential $P$ given in Eq.(\ref{P_mod}), 
$D_\phi P$ and ${\cal D}_\phi D_\phi P$ are of the form:
\begin{eqnarray}
D_\phi P 
&=& {\rm e}^{-\frac{\kappa^2}{2}\phi^2} 
\left[\kappa^2 (\phi^* - \phi) 
\frac{\Lambda^3}{g^2} {\rm sin} \frac{g}{\Lambda} \phi 
+  
\frac{\Lambda^2}{g} {\rm cos}\frac{g}{\Lambda}\phi 
\right],\\
{\cal D}_\phi D_\phi P 
&=& 2 \kappa^2 (\phi^*-\phi) \partial_\phi P 
- \left(\kappa^2 + \kappa^4 (\phi^2 - \phi^{*2}) 
+ \frac{g^2}{\Lambda^2}\right)P, \\
&=&{\rm e}^{-\frac{\kappa^2}{2}\phi^2}
\left[ -\left(\kappa^2 + \frac{g^2}{\Lambda^2} 
- \kappa^4 (\phi - \phi^*)^2\right)
\frac{\Lambda^3}{g^2}
{\rm sin}\frac{g}{\Lambda} \phi 
+ 2 \kappa^2 (\phi^* - \phi) 
\frac{\Lambda^2}{g} {\rm cos}\frac{g}{\Lambda} \phi 
\right].
\label{eq:DDP}
\end{eqnarray}

\subsection{Gravitino}

In this subsection, we will explore a massless gravitino which is 
a superpartner of the massless localized graviton under the 
SUGRA transformation with the conserved 
Killing spinor (\ref{eq:Killingspinor}). 
Before studying equations of motion for gravitino, 
we will supertransform the wave function of the localized 
massless graviton to find conditions that the physical 
gravitino should satisfy. 
Let us focus on SUGRA transformation law for vierbein 
in Eq.(\ref{eq:SUGRAtransf}), 
\begin{equation}
\delta_\zeta e_m{^{\underline{a}}}
= i \kappa \left( \zeta \sigma^{\underline{a}} \bar{\psi}_m 
+ \bar{\zeta} \bar{\sigma}^{\underline{a}} \psi_m \right). 
\label{vtrf}
\end{equation}
The preserved SUSY along the Killing spinor 
$\zeta{(K)}$ in Eq.(\ref{eq:Killingspinor}) 
with $\theta=\pi/2$ is given by 
\begin{eqnarray}
\zeta^\alpha{(K)}
&=& -i \bar{\zeta}_{\dot{\alpha}}(K)
\bar{\sigma}^{2 \dot{\alpha} \alpha} 
= i {\rm e}^{A/2} [\epsilon_2, -\epsilon_1], \\
\bar{\zeta}_{\dot{\alpha}}(K)
&=& -i \zeta^\alpha{(K)} \sigma^2_{\alpha \dot{\alpha}} 
= i{\rm e}^{A/2} [-\epsilon_1, -\epsilon_2]. 
\label{eq:Killing-spinor}
\end{eqnarray}
Denoting the fluctuations $h_{mn}$ of the metric 
around the background spacetime metric 
$g_{mn}^{\rm background}\equiv diag({\rm e}^{2A}\eta_{\mu\nu}, 1)$ as 
$g_{mn}=g_{mn}^{\rm background} + h_{mn}$, 
the following linearized 3D SUGRA transformations 
with the Killing spinor $\zeta(K)$ are obtained for 
the metric fluctuations 
$\delta h_{mn}
=\delta(e_m{}^{\underline{a}} 
e_{n \underline{a}})
=\delta e_m{}^{\underline{a}} 
 e_{n \underline{a}}
+ e_{m\underline{a}} 
\delta e_{n}{}^{\underline{a}}$ 
\begin{eqnarray}
\label{3dmntrf}
\delta_{\zeta(K)} h_{\mu \nu} &=& i \kappa {\rm e}^A \zeta(K) 
(\sigma_{\underline{\mu}} \bar{\psi}_\nu -i \sigma^{\underline{2}} 
\bar{\sigma}_{\underline{\mu}} \psi_\nu + \sigma_{\underline{\nu}} 
\bar{\psi}_\mu -i \sigma^{\underline{2}} \bar{\sigma}_{\underline{\nu}} \psi_\mu), \\
\label{3d2mtrf}
\delta_{\zeta(K)} h_{2 \mu} &=& i \kappa \zeta(K) 
\left\{ \sigma_{\underline{2}} \bar{\psi}_\mu + i \psi_\mu 
+ (\sigma_{\underline{\mu}} \bar{\psi}_2 -i\sigma^{\underline{2}} 
\bar{\sigma}_{\underline{\mu}} \psi_2) {\rm e}^A \right\}, \\
\label{3d22trf}
\delta_{\zeta(K)} h_{22} &=& 2i \kappa \zeta(K) (\sigma_2 \bar{\psi}_2 
+i \psi_2). 
\end{eqnarray}

In sect.\ref{sc:tensot-perturb}, 
we have imposed the gauge fixing condition (Newton Gauge) for graviton 
\begin{equation}
h_{22}= - \frac{1}{3} {\rm e}^{-2A}\eta^{\mu\nu}h_{\mu\nu} \equiv - \frac{1}{3}h, 
\label{eq:gauge-fix-graviton1} 
\end{equation}
\begin{equation}
h_{2\mu}=0.
\label{eq:gauge-fix-graviton2} 
\end{equation}
We can algebraically decompose $h_{\mu\nu}$ into traceless part 
${\rm e}^{2A}h_{\mu\nu}^{TT}$ and trace part $h$. 
We have found that 
the localized graviton zero mode is contained in the traceless part 
\begin{eqnarray}
\eta^{\mu \nu} h_{\mu \nu} = 0.
\label{eq:traceless}
\end{eqnarray}
Equation of motion shows that the localized graviton zero mode 
also satisfies the
transverse condition:
\begin{eqnarray}
\eta^{\lambda \mu} \partial_\lambda h_{\mu \nu} = 0.
\label{eq:transverse}
\end{eqnarray}
The matter fermion of course do not have 
the graviton zero mode : $\varphi=0$. 

It is useful to decompose Weyl spinors in four dimensions into 
two 2-component Majorana spinors in three dimensions. 
For instance gravitinos $\psi_m$ are decomposed into 
two 2-component Majorana spinor-vectors $\psi_m^{(1)}$ 
and  $\psi_m^{(2)}$ (real and imaginary part of 
the Weyl spinor-vector) 
\begin{eqnarray}
\label{majo1}
\psi_{m\alpha}^{(1)} \equiv \psi_{m\alpha} 
- i \sigma^2_{\alpha\dot\alpha} \bar{\psi}_m^{\dot\alpha}
=-i\sigma^2_{\alpha\dot\alpha}\bar\psi_m^{(1)\dot\alpha},
\end{eqnarray}
\begin{eqnarray}
\label{majo2}
\psi^{(2)}_{m\alpha} \equiv \psi_{m\alpha} 
+ i \sigma^2_{\alpha\dot\alpha} \bar{\psi}_m^{\dot\alpha}
=i\sigma^2_{\alpha\dot\alpha}\bar\psi_m^{(2)\dot\alpha}. 
\end{eqnarray}
Similarly to the traceless and trace part decomposition of graviton 
(symmetric tensor), 
gravitino (vector-spinor) can also be algebraically decomposed 
into its 
traceless part $\psi_\mu^T$ and trace part $\bar \psi$ as 
\begin{equation}
\psi_\mu=\psi_\mu^T 
- {1 \over 3}\sigma_{\underline{\mu}}\bar\psi, 
\qquad 
\bar \sigma^{\underline{\mu}}\psi_\mu^T=0, 
\qquad
\bar \sigma^{\underline{\mu}}\psi_\mu=\bar\psi.
\end{equation}

Let us make a SUGRA transformations of the physical state 
conditions (\ref{eq:gauge-fix-graviton1})--(\ref{eq:transverse}) 
for gravitons with the  conserved Killing spinor $\zeta(K)$ in 
 Eq.(\ref{eq:Killing-spinor}). 
The SUGRA transformations with $\zeta(K)$ of 
Eqs.(\ref{eq:gauge-fix-graviton1}), (\ref{eq:gauge-fix-graviton2})
(\ref{eq:traceless}) give 
\begin{eqnarray}
0 = \delta_{\zeta(K)} \left( h_{22} + \frac{1}{3} h\right) 
= 2i \kappa \zeta(K) \left(
\sigma_{\underline{2}} 
\bar{\psi}_2^{(1)}
+ \frac{1}{3} {\rm e}^{-A}\sigma^{\underline{\mu}}
\bar\psi_\mu^{(2)}
\right), 
\label{eq:h22gauge-trans}
\end{eqnarray}
\begin{eqnarray}
0 &=& \delta_{\zeta(K)} h_{2\mu} 
= i \kappa \zeta 
\left(
\sigma_{\underline{2}} \bar{\psi}_\mu^{(1)} 
+ {\rm e}^A 
\sigma_{\underline{\mu}} \bar{\psi}_2^{(2)} 
\right).
\label{eq:gauge-fix-graviton2(2)}
\end{eqnarray}
\begin{eqnarray}
0 &=& \delta_{\zeta(K)} \eta^{\mu \nu} h_{\mu \nu} 
=
 2i\kappa {\rm e}^{-A} \zeta 
\sigma^{\underline{\mu}} \bar{\psi}_{\mu}^{(2)}. 
\label{eq:traceless-gauge-tr}
\end{eqnarray}
These result suggest the most natural gauge fixing condition 
for local gauge SUGRA transformations 
\begin{eqnarray}
\psi_2 =0, 
\label{eq:psi2-gauge}
\end{eqnarray}
which can 
always be chosen. 
Then, the above gauge fixing conditions 
(\ref{eq:h22gauge-trans})--(\ref{eq:traceless-gauge-tr}) 
are translated as $\bar\psi_\mu^{(1)} = 0$ and 
the traceless condition for $\psi^{(2)}$ 
\begin{eqnarray}
\sigma^{\underline{\mu}}\bar\psi_\mu = 0.
\label{gravitino3}
\end{eqnarray}
Therefore we expect\footnote{
One should have in mind that it is desirable to choose 
a gauge fixing condition for SUGRA transformations 
to become a supertransformation under $\zeta(K)$ of the gauge fixing 
condition for general coordinate transformations. 
However, it may not be logically mandatory. 
} 
that the localized massless 
gravitino should be contained in the traceless part of $\psi_\mu^{(2)}$. 
The SUGRA transformation of the remaining condition 
(\ref{eq:transverse}) gives
the transverse condition for the $\psi_\mu^{(2)}$.
Similarly to the graviton case, the localized 
gravitino should not have matter component 
\begin{equation}
\chi=0. 
\label{eq:no-matter-fermion}
\end{equation}

Let us now examine the 
equations of motion for gravitino $\psi_\mu$ coupled with 
the matter fermion $\chi$, which are obtained 
by varying the action (\ref{eq:fermion-lag}). 
If we impose the conditions (\ref{eq:psi2-gauge}), 
(\ref{gravitino3}), and (\ref{eq:no-matter-fermion}) on 
the gravitino equations of motion, we obtain 
\begin{eqnarray}
0 &=& 
{\rm e}^{-3A} 
\vep^{\underline{\mu 2 \rho \nu}} \bsig_{\underline{2}} 
\partial_\rho \psi_\nu 
+{\rm e}^{-2A} 
\left(
- \half \dot{A} \vep^{\underline{\mu 2 \rho \nu}} 
\bsig_{\underline{\rho}} \psi_\nu 
+ \vep^{\underline{\mu \rho 2 \nu}} 
\bsig_{\underline{\rho}} \partial_2 \psi_\nu 
 - \kappa^2{\rm e}^{{\kappa^2 \over 2}\phi^*\phi}
 P \bar\sigma^{\mu\nu}\bpsi_\nu 
 \right) \nonumber \\
 &=& 
{\rm e}^{-3A} 
\vep^{\underline{\mu 2 \rho \nu}} \bsig_{\underline{2}} 
\partial_\rho \psi_\nu 
+{\rm e}^{-2A} 
\left[
-   \vep^{\underline{\mu 2 \rho \nu}} 
\bsig_{\underline{\rho}} 
\left(\partial_2 +\half \dot{A}\right) \psi_\nu 
 + \dot{A}  \eta^{\mu\nu}\bpsi_\nu 
 \right],   
\end{eqnarray}
where we have used 
the BPS equation (\ref{BPS_A}) 
for background fields. 
Possible zero mode should give 
a vanishing eigenvalue for the operator 
in the parenthesis : 
\begin{eqnarray}
0 &=& 
-   \vep^{\underline{\mu 2 \rho \nu}} 
\bsig_{\underline{\rho}} 
\left(\partial_2 +\half \dot{A}\right) \psi_\nu 
 + \dot{A}  \eta^{\mu\nu}\bpsi_\nu 
=-i \eta^{\mu\nu} \bar{\sigma}^{\underline{2}} 
\left(\partial_2 + {\dot{A}\over 2}\right) 
\psi_\nu 
 + \dot{A} \eta^{\mu\nu}\bpsi_\nu 
\label{eom1}
\end{eqnarray}
where 
$\vep^{\underline{\mu \rho 2 \nu}} 
\bsig_{\underline{\rho}}= i
(\bsig^{\underline{\mu}} 
\sig^{\underline{2}} \bsig^{\underline{\nu}} 
- \eta^{\underline{\mu \nu}} \bsig^{\underline{2}})$ is used 
in the second equality. 
In terms of the 2-component Majorana spinors (\ref{majo1}) 
and (\ref{majo2}), we obtain 
\begin{eqnarray}
\left(\partial_2 + \frac{3}{2}\dot A\right)\bar\psi^{(1)TT}_\mu 
+ \left(-\partial_2 + \frac{1}{2}\dot A\right)\bar\psi^{(2)TT}_\mu = 0.
\end{eqnarray}
Since $\psi^{(1)}_\mu = 0$,
we obtain
\begin{eqnarray}
\left(-\partial_2 + \frac{1}{2}\dot A\right)\bar\psi^{(2)TT}_\mu = 0.
\end{eqnarray}
Therefore, we find the gravitino zero mode 
in the transverse traceless part of the 2-component Majorana 
vector-spinor $\bar\psi^{(2)TT}_\mu$
with the wave function
\begin{eqnarray}
\bar\psi^{(2)TT}_\mu(y) = {\rm e}^{\frac{A(y)}{2}}.
\end{eqnarray}

Now we see that the localized massless gravitino wave 
function is in precise agreement with that expected 
from the preserved 
SUGRA transformation with the Killing spinor $\zeta(K)$ : 
\begin{eqnarray}
e_{\mu}{^{\underline{a}}} \sim \zeta(K) \sigma^{\underline{a}}\psi_{\mu}.
\end{eqnarray}
Since the wave function of the graviton and 
the Killing spinor are 
$e_\mu{^{\underline{a}}} \sim {\rm e}^{A}$ 
and $\zeta(K) = {\rm e}^{\frac{A}{2}}$, 
we find $\psi_\mu \sim {\rm e}^{\frac{A}{2}}$.

\subsection{Matter Fermion}

In this subsection, we study the fluctuation of 
matter fermion. 
By varying the Lagrangian (\ref{eq:fermion-lag}) 
with respect to $\chi$, we obtain the equation of motion 
for matter fermion $\chi$. 
Using the gauge choice 
$\psi_2 = \bpsi_2 = 0$ to the equation of motion, we find 
\begin{eqnarray}
0 &=& 
-i{\rm e}^{A} \bar\sig^{\underline{\mu}} 
\partial_\mu {\chi} 
-\frac{3}{2}i \dot{A}  \bsig_2 {\chi}
-i  \bsig^{\underline{2}} \partial_2 {\chi} 
-{\rm e}^{\frac{\kappa^2}{2}\phi^* \phi} 
{\cal D}_{\phi^*} D_{\phi^*} P^* \bar{{\chi}} 
\nonumber \\
&&- \frac{\kappa}{\sqrt{2}}  \dot{\phi} 
\bsig^{\underline{\mu}} \sig^{\underline{2}} \bpsi_\mu 
-{\rm e}^{\frac{\kappa^2}{2}\phi^* \phi} 
\frac{i\kappa}{\sqrt{2}}  D_{\phi^*} P^* 
\bsig^{\underline{\mu}} \psi_\mu 
. 
\end{eqnarray}
The second line gives 
the mixing term between the trace part of 
gravitino $\bar\psi=\bar\sigma^\mu\psi_\mu$ 
and the matter fermion $\chi$. 
Using the BPS equation (\ref{BPS_phi}), the mixing term 
can be rewritten as 
\begin{eqnarray}
- \frac{\kappa}{\sqrt{2}}  \dot{\phi} 
\bsig^{\underline{\mu}} \sig^{\underline{2}} \bpsi_\mu 
-\frac{i\kappa}{\sqrt{2}} \dot{\phi} 
\bsig^{\underline{\mu}} \psi_\mu
= 
-\frac{i\kappa}{\sqrt{2}} \dot{\phi} 
\bsig^{\underline{\mu}}\left(\psi_\mu 
- i\sig^{\underline{2}} \bar \psi_\mu\right)
= 
-\frac{i\kappa}{\sqrt{2}} \dot{\phi} 
\bsig^{\underline{\mu}}\psi_\mu^{(1)} ,
\end{eqnarray}
where we used the 2-component Majorana spinor notation 
defined in Eqs.(\ref{majo1}), (\ref{majo2}). 
Since the mixing occurs only with $\psi_\mu^{(1)}$, 
it is also useful to decompose matter fermions 
into 2-component Majorana spinors, similarly to 
Eqs.(\ref{majo1}), (\ref{majo2}). 
Then the matter equation of motion is decomposed into 
two parts with opposite transformation property under 
the charge conjugation $\sigma^2$ 
\begin{equation}
-i{\rm e}^{A} \bar\sig^{\underline{\mu}} 
\partial_\mu {\chi}^{(2)} 
= i  \bsig^{\underline{2}} \left[
\partial_2 +\frac{3}{2} \dot{A}
-{\rm e}^{\frac{\kappa^2}{2}\phi^* \phi} 
{\cal D}_{\phi^*} D_{\phi^*} P^* \right] 
{\chi}^{(1)} 
,
\end{equation}
\begin{equation}
-i{\rm e}^{A} \bar\sig^{\underline{\mu}} 
\partial_\mu {\chi}^{(1)} 
=i  \bsig^{\underline{2}} \left[
\partial_2 +\frac{3}{2} \dot{A}
+{\rm e}^{\frac{\kappa^2}{2}\phi^* \phi} 
{\cal D}_{\phi^*} D_{\phi^*} P^* \right] 
{\chi}^{(2)} 
-i\sqrt2 \kappa \dot{\phi} \bar \sigma^{\underline{\mu}}
\psi_\mu^{(1)}
. 
\end{equation}
It is now clear that we have a zero mode 
consisting of purely $\chi^{(1)}$ : 
\begin{equation}
-i{\rm e}^{A} \bar\sig^{\underline{\mu}} 
\partial_\mu {\chi}^{(1)} 
=0
,
\qquad \chi^{(2)}=0, 
\qquad \psi_\mu=0
\end{equation}
The zero mode wave function for matter fermion is given by 
\begin{equation}
\left[
\partial_2 +\frac{3}{2} \dot{A}
-{\rm e}^{\frac{\kappa^2}{2}\phi^* \phi} 
{\cal D}_{\phi^*} D_{\phi^*} P^* \right] 
{\chi}^{(1)} =0 , 
\end{equation}
whose solution is given by 
\begin{equation}
{\chi}^{(1)}_0 \sim {\rm e}^{-3A/2} {\rm exp} 
\left[ \int dy \ {\rm e}^{\frac{\kappa^2}{2}\phi^* \phi} 
{\cal D}_\phi 
D_\phi P \right], 
\label{matterzero}
\end{equation}
Using (\ref{eq:DDP}), 
the integral in (\ref{matterzero}) in BPS case 
reads 
\begin{eqnarray}
\label{intm32}
\int dy \ {\rm e}^{\frac{\kappa^2}{2}\phi^* \phi} 
{\cal D}_\phi D_\phi P 
&=& 
-(\kappa^2 + \frac{g^2}{\Lambda^2}) \frac{\Lambda^3}{g^2} 
\int dy \ {\rm sin} \frac{g}{\Lambda} \phi
=-(1+\kappa^2 \frac{\Lambda^2}{g^2} ) \Lambda 
\int dy \ {\rm tanh}(\Lambda y) \nonumber \\
&=& 
-(1+\kappa^2 \frac{\Lambda^2}{g^2} ) 
{\rm log}({\rm cosh}(\Lambda y)) .
\end{eqnarray}
In the second equality, 
$P={\rm e}^{-\kappa^2 \phi^2/2}
\frac{\Lambda^3}{g^2}{\rm sin}(g\phi/\Lambda)$ 
in Eq.(\ref{P_mod}) is substituted and  
$\phi = \phi^*$ is taken into account. 
In the last equality, 
the BPS solution $\phi = \frac{\Lambda}{g}
\left(
2{\rm tan}^{-1}e^{\Lambda(y-y_0)} - \frac{\pi}{2}
\right)$ in Eq.(\ref{BPS_sol}) 
with $n=0$ is considered. 
Then the zero mode wave function of matter fermion 
 $\chi^{(1)}$ is 
given by 
\begin{eqnarray}
\label{mf01}
\chi^{(1)}_0 &\sim& {\rm e}^{-3A/2} 
\left[{\rm cosh}(\Lambda y)
\right]^{-\left[1 + \kappa^2 \frac{\Lambda^2}{g^2} \right]}.  
\end{eqnarray}

In the weak gravity limit, the zero mode of matter fermion 
reduces to the Nambu-Goldstone fermion associated with 
the spontaneously broken SUSY \cite{MSSS2} 
\begin{eqnarray}
\label{mf01-1}
\chi^{(1)}_0 
 \to 
\frac{1}{{\rm cosh}(\Lambda y)}, \qquad 
\kappa \to 0 . 
\end{eqnarray}
As expected 
in the global SUSY limit, 
the wave function is localized at the wall where two 
out of four SUSY are broken. 
Let us note, however, that this zero mode of matter fermion 
should be unphysical except at $\kappa =0$ limit. 
For any finite values of $\kappa$, it should be possible 
to gauge away this zero mode, precisely analogously to 
the zero modes $\hat{B}^{(+)}$ in Eq.(\ref{zero_scalar}) 
in the matter scalar sector in sect.\ref{sc:active-scalar}. 
In fact we can see that the $A$ dependence (warp factor) 
of the zero modes of active scalar 
$\hat{B}^{(+)}(x)$ and the matter fermion 
$\chi^{(1)}_0(x)$ agrees with the surviving 
SUSY transformation generated by the Killing 
spinor $\zeta(K)$, and will form a supermultiplet 
under the surviving SUGRA, 
since 
$\phi_{{\rm R}}^{(0)}(y) \sim e^{-A}, \chi^{(1)}_0(y) 
\sim {\rm e}^{-3A/2}$ 
and $\zeta(y) \sim {\rm e}^{A/2}$ 
\begin{eqnarray}
\delta_\zeta \phi(x,y)_{{\rm R}} 
= \sqrt{2} \zeta(x,y) \chi^{(1)}(x,y). 
\end{eqnarray}

On the other hand, ${\chi}^{(2)}$ should contain 
another Nambu-Goldstone fermion corresponding to 
the SUSY charges broken by the negative tension 
brane, if we consider non-BPS multi-wall configurations. 
Noting that the mixing term is suppressed 
by the Planck scale $M_P$, 
the zero mode equation of motion for ${\chi}^{(2)}$ 
in weak gravity limit $\kappa \to 0$ 
is given by 
\begin{eqnarray}
0 &=& \partial_2 {\chi}^{(2)} 
+ {\rm e}^{\frac{\kappa^2}{2}\phi^* \phi} 
{\cal D}_\phi D_\phi P {\chi}^{(2)}, \\
&\to& \partial_2 {\chi}^{(2)} 
- \Lambda {\rm sin}\frac{g}{\Lambda} 
\phi~{\chi}^{(2)} 
= \partial_2 {\chi}^{(2)} - \Lambda {\rm tanh}(\Lambda y) 
{\chi}^{(2)}, 
\quad (\kappa \to 0) 
\end{eqnarray}
where BPS solution 
$\phi = \frac{\Lambda}{g}
\left(
2{\rm tan}^{-1}e^{\Lambda(y-y_0)} - \frac{\pi}{2}
\right)$
is substituted in the last equality, 
thus the zero mode wave function becomes
\begin{equation}
{\chi}^{(2)}_0 \to {\rm cosh}(\Lambda y)~(\kappa \to 0), 
\end{equation}
which is not normalizable and hence unphysical even in 
the limit of $\kappa\rightarrow 0$. 
We know from the exact solution of the non-BPS two-wall 
solution\cite{MSSS2}, 
that this wave function results when taking 
the limit of large radius to obtain the BPS solution. 
In that limit, SUSY broken on the second wall at $y=\pi R$ 
is restored and the corresponding Nambu-Goldstone fermion, 
which was localized on the second brane, 
becomes non-normalizable and unphysical. 
This is precisely our zero mode 
wave function $\chi^{(2)}_0$.

%%%%%%%%%%%%%%%%%%%%%%%%%%%%%%%%%%%%%%%%%%%%%%%%%%%%%%%%%%%%%%%%%%%%%%%%%%%%%%%
\section*{Acknowledgements}
We thank Daisuke Ida, Kazuya Koyama, Tetsuya Shiromizu, and 
Takahiro Tanaka for useful discussions in several occasions. 
One of the authors (M.E.) gratefully acknowledges support from 
the Iwanami Fujukai Foundation.
This work is supported in part by Grant-in-Aid for Scientific 
Research from the Ministry of Education, Culture, Sports, 
Science and 
Technology, Japan No.13640269 (NS) and 
by Special Postdoctoral Researchers Program at RIKEN (NM). 

%%%%%%%%%%%%%%%%%%%%%%%%%%%%%%%%%%%%%%%%%%%%%%%%%%%%%%%%%%%%%%%%%%%%%%%%%%%%%%%%

\renewcommand{\thesubsection}
{\thesection.\arabic{subsection}}

\appendix

\section{Appendix A}
In this appendix we show the gauge fixing to the Newton gauge.
The most general fluctuation around the background metric
(\ref{warped_metric}) takes the form : 
\begin{eqnarray}
ds^2 &=& {\rm e}^{2A}\left(\eta_{\mu\nu} + h^{\rm T}_{\mu\nu} 
+ 2\eta_{\mu\nu}B\right)dx^\mu dx^\nu
+ 2 f_{\mu}\ dx^\mu dy + \left(1 - 2C\right)dy^2,
\end{eqnarray}
where the trace part of the $(\mu,\nu)$ component of fluctuations 
is denoted as $2\eta_{\mu\nu}B$ and the
traceless part is denoted as $h^{\rm T}_{\mu\nu}$. $f_\mu$ denotes the
fluctuation of $(\mu,2)$ component and $-2C$ denotes the fluctuation of
$(2,2)$ component.

After a tedious calculation we find the linearized Ricci tensors : 
\begin{eqnarray}
R^{(1)}_{\mu\nu}
&\!\!\!=&\!\!\! 
\frac{1}{2}{\rm e}^{2A}\left({\rm e}^{-2A}\square_{D-1} 
+ \partial_y^2
+ (D-1)\dot{A}\partial_y 
+ 2(D-1)\dot{A}^2 
+ 2\ddot{A}
\right)
h^{\rm T}_{\mu\nu}
- h^{\rm T}_{(\mu\rho,\nu)}{^{,\rho}}\nonumber\\
&\!\!\!&\!\!\! 
+ \eta_{\mu\nu}{\rm e}^{2A}\left({\rm e}^{-2A}\square_{D-1} 
+ \partial_y^2 
+ 2(D-1)\left(\dot{A}\partial_y
+ \dot{A}^2 \right) 
+ 2\ddot{A}
\right)B
+ (D-3)B_{,\mu,\nu}\nonumber\\
&\!\!\!&\!\!\!
- \eta_{\mu\nu}\dot{A} 
f_\rho{^{,\rho}}
- \left(\partial_y 
+ (D-3)\dot{A}
\right)f_{(\mu,\nu)}
+ \eta_{\mu\nu}{\rm e}^{2A}\left(\dot{A}\partial_y 
+ 2\ddot{A} 
+ 2(D-1)\dot{A}^2
\right)C
- C_{,\mu,\nu},
\\\ \nonumber\\
%%%%%%%%%%%%%%%%%%%%%%%%%%%%%%%%%%%%%%%%%%%%%%%%%%%%%%%%%%%%%%%%%%%%%%%%%%%%%%%%%%%%%%%%%%%%%%%%%%%%%%%%%%%%
R^{(1)}_{\mu2} 
&\!\!\!=&\!\!\! 
- \frac{1}{2}\left(-{\rm e}^{-2A}\square_{D-1} 
- 2(D-1)\dot{A}^2 
- 2\ddot{A}
\right)f_{\mu}
- \frac{1}{2}{\rm e}^{-2A}f_{\rho,\mu}{^{,\rho}}\nonumber\\
&\!\!\!&\!\!\!
+ (D-2)\dot{A}
C_{,\mu}
+ (D-2)\partial_y
B_{,\mu}
- \frac{1}{2}\partial_y
h^{\rm T}_{\mu\rho}{^{,\rho}},
\\\ \nonumber\\
%%%%%%%%%%%%%%%%%%%%%%%%%%%%%%%%%%%%%%%%%%%%%%%%%%%%%%%%%%%%%%%%%%%%%%%%%%%%%%%%%%%%%%%%%%%%%%%%%%%%%%%%%%%%
R^{(1)}_{22}
&\!\!\!=&\!\!\! 
(D-1)\left(\partial_y^2
+ 2\dot{A}\partial_y
\right)B
- {\rm e}^{-2A}\partial_y
f_{\rho}{^{,\rho}}
- \left({\rm e}^{-2A}\square_{D-1} - (D-1)\dot{A}\partial_y
\right)C,
\end{eqnarray}
where we define $B_{,\mu} = \partial_\mu B$, 
$h^{\rm T}_{(\mu\rho,\nu)} = \dfrac{1}{2}\left(h^{\rm T}_{\mu\rho,\nu} +
h^{\rm T}_{\nu\rho,\mu}\right)$, $f_\rho{^{,\rho}} =
\eta^{\rho\lambda}\partial_\lambda f_\rho$ and $\square_{D-1} =
\eta^{\rho\lambda}\partial_\rho\partial_\lambda$. 
We also find the linearized energy momentum tensor as follows : 
\begin{eqnarray}
\tilde{T}^{(1)}_{\mu\nu}
&\!\!\!=&\!\!\! {2 \over D-2}{\rm e}^{2A}\left[
V_{\rm R}h^{\rm T}_{\mu\nu} + \eta_{\mu\nu}\left(2V_{\rm R}B 
+ \frac{dV_{\rm R}}{d\phi_{\rm R}}\varphi_{\rm R}\right)\right],\\
\tilde{T}^{(1)}_{\mu2}
&\!\!\!=&\!\!\! 2\dot\phi_{\rm R}
\varphi_{{\rm R},\mu} + {2 \over D-2}V_{\rm R}f_\mu,\\
\tilde{T}^{(1)}_{22} 
&\!\!\!=&\!\!\! 
4 \dot\phi_{\rm R}\partial_y
\varphi_{\rm R} 
+ {2 \over D-2}\frac{dV_{\rm R}}{d\phi_{\rm R}}\varphi_{\rm R} 
- {4 \over D-2}V_{\rm R}C,
\end{eqnarray}
where $\varphi_{\rm R}$ is the fluctuation around the background 
active scalar 
field $\phi_{\rm R}$. 
Notice that the fluctuation $\varphi_{\rm I}$
about the background configuration for the imaginary part 
$\phi_{\rm I}$ 
decouples from any other fields in linear order of the fluctuations. 
We can 
obtain the linearized Einstein equations by plugging these into 
$R^{(1)}_{mn} = -\kappa^2\tilde{T}^{(1)}_{mn}$.

The above results are the most general in the sense 
that we do not fix
any gauge for the fluctuations. 
As a next step, we wish to fix the gauge that 
simplifies the linearized equations. 
The ``Newton'' gauge is 
known as a candidate of such a gauge 
\cite{TanakaMontes, CsabaCsaki}. 
The gauge transformation laws for 
the fluctuations are of the form : 
\begin{eqnarray}
\delta h^{\rm T}_{\mu\nu} 
= 
- \hat\xi_{(\mu,\nu)}
+ \dfrac{2}{D-1}\eta_{\mu\nu}\hat\xi_{\rho}{^{,\rho}},\quad
\delta B 
= 
- \dot{A}
\xi_2 - \dfrac{1}{D-1}\hat\xi_{\rho}{^{,\rho}},
\nonumber\\
\delta f_{\mu} 
= 
- {\rm e}^{2A}\partial_y
\hat{\xi}_{\mu} - \xi_{2,\mu},\quad
\delta C 
= 
\partial_y
\xi_{2},\quad
\delta\varphi_{\rm R} = - \dot\phi_{\rm R}
\xi_2,
\end{eqnarray}
where $\xi_m$ is an infinitesimal coordinate 
transformation parameter
$\delta x_m \equiv \xi_m$ and 
$\hat\xi_\mu \equiv {\rm e}^{-2A}\xi_\mu$.
Using these four gauge freedom, we fix 
$f_\mu = 0$ and $(D-3)B = C$. 
The 
residual gauge transformation should satisfy
\begin{eqnarray}
\partial_y
\hat\xi_\mu + {\rm e}^{-2A}\xi_{2,\mu}=0,\quad
\left(\partial_y
+(D-3)\dot{A}
\right)\xi_2 =
-\dfrac{D-3}{D-1}\hat\xi_\rho{^{,\rho}}.
\label{residual}
\end{eqnarray}
In this gauge the linearized Einstein equations 
take the form : 
\begin{alignat}{2}
\!\!\!&\frac{1}{2}{\rm e}^{2A}
\left({\rm e}^{-2A}\square_{D-1} +
\partial_y^2
+ (D-1)\dot{A}\partial_y
+ 2(D-1)\dot{A}^2
+ 2\ddot{A}
\right)h^{\rm T}_{\mu\nu} 
- h^{\rm T}_{(\mu\rho,\nu)}{^{,\rho}}
+ \frac{1}{D-1}\eta_{\mu\nu}h^{\rm T}_{\rho\lambda}
{^{,\rho,\lambda}} 
\nonumber \\
\!\!\!&
= - {2 \over D-2}\kappa^2 {\rm e}^{2A} 
V_{\rm R}h^{\rm T}_{\mu\nu},
\label{1st_gauge_1}
\\
\!\!\!&{\rm e}^{2A}\left({\rm e}^{-2A}\square_{D-1} 
+ \partial_y^2
+ (3D-5)\dot{A}\partial_y
+ 2(D-2)\left((D-1)\dot{A}^2
+ \ddot{A}\right)
\right)B
- \frac{1}{D-1}h^{\rm T}_{\rho\lambda}{^{,\rho,\lambda}}
\nonumber \\
\!\!\!&= -{2 \over D-2}\kappa^2{\rm e}^{2A}
\left(2V_{\rm R}B + \frac{dV_{\rm R}}
{d\phi_{\rm R}}\varphi_{\rm R}\right),\label{1st_gauge_2}\\
\!\!\!&(D-2)\left(\partial_y + (D-3)\dot{A}
\right)B_{,\mu}
- \frac{1}{2}\partial_y
h^{\rm T}_{\mu\rho}{^{,\rho}}
= -2\kappa^2\dot\phi_{\rm R}
\varphi_{{\rm R},\mu},\label{1st_gauge_3'}
\\
\!\!\!&\left(-(D-3){\rm e}^{-2A}\square_{D-1} +
(D-1)\partial_y^2
+ (D-1)^2\dot{A}\partial_y
\right)B
\nonumber \\
\!\!\!&
= -2\kappa^2\left(
2\dot\phi_{\rm R}\partial_y
\varphi_{\rm R}
+ \frac{1}{D-2}\frac{dV_{\rm R}}{d\phi_{\rm R}}
\varphi_{\rm R} - {2(D-3) \over D-2}V_{\rm R}B
\right),\label{1st_gauge_4'}
\end{alignat}
where the Eq.(\ref{1st_gauge_1}) is the traceless 
part of the $(\mu,\nu)$ 
component whereas the Eq.(\ref{1st_gauge_2}) is 
the trace part of 
it. Denoting $D(x,y) = (D-2)\left(\partial_y+(D-3)\dot{A}\right)B
+ 2\kappa^2\dot\phi_{\rm R}
\varphi_{\rm R}$, the
Eq.(\ref{1st_gauge_3'}) is rewritten as 
\begin{eqnarray}
D_{,\mu}(x,y) = 
\dfrac{1}{2}\partial_y
h^{\rm T}_{\mu\rho}{^{,\rho}}.\label{1st_gauge_3}
\end{eqnarray}
Summing the background Einstein equation 
Eq.(\ref{1st_gauge_2}) multiplied by 
$(D-3){\rm e}^{-2A}$ and 
Eq.(\ref{1st_gauge_4'}) gives 
$\left(\partial_y + (D-1)\dot{A}\right)D(x,y)
= \dfrac{D-3}{2(D-1)}{\rm e}^{-2A}
h^{\rm T}_{\rho\lambda}{^{,\rho,\lambda}}$. 
Then we find 
\begin{eqnarray}
\displaystyle D(x,y) = E(x)\ {\rm e}^{-(D-1)A(y)} + 
\dfrac{D-3}{2(D-1)}{\rm e}^{-(D-1)A}\int dy\ 
{\rm e}^{(D-3)A}
h^{\rm T}_{\rho\lambda}{^{,\rho,\lambda}}
\label{1st_gauge_4}
\end{eqnarray}
where $E(x)$ is an arbitrary function of $x$. 
At this stage, the
linearized Einstein equations are Eq.(\ref{1st_gauge_1}),
(\ref{1st_gauge_2}), (\ref{1st_gauge_3}) 
and (\ref{1st_gauge_4}).

Next we attempt to eliminate the longitudinal mode of 
$h^{\rm T}_{\mu\nu}$ by using the residual gauge freedom. 
That is, we wish to
set $v_\mu \equiv h^{\rm T}_{\mu\rho}{^{,\rho}} = 0$. 
For that purpose we first derive the equations of 
motion for $v_\mu$.
Taking a divergence of Eq.(\ref{1st_gauge_1}), we find
\begin{eqnarray}
\left(\partial_y
+ (D-1)\dot{A}
\right)\partial_yv_\mu 
= \frac{D-3}{D-1}
{\rm e}^{-2A}v_\rho{^{,\rho}}{_{,\mu}}.\label{eq_v}
\end{eqnarray}
This equation can be solved as follows :  
i) taking divergence, we can determine
$v_\rho{^{,\rho}}$, ii) regarding the solution 
$v_\rho{^{,\rho}}$ as a source,
we can determine $v_\mu$.
The gauge transformation law of $v_\mu$ takes the form
\begin{eqnarray}
\delta v_\mu 
= - \square_{D-1} \hat\xi_\mu - 
\frac{D-3}{D-1}\hat\xi_{\rho,\mu}{^{,\rho}}.\label{delta_v}
\end{eqnarray}
We want to set $0 = v_\mu + \delta v_\mu$ by using the gauge
transformation (\ref{delta_v}) whose $\hat\xi_\mu$ 
satisfies the
condition (\ref{residual}) for the residual gauge 
transformation.
Notice that the equations for $v_\mu$ and $\delta v_\mu$ 
are identical second
order differential equations since the gauge 
transformation law
consistent with the gauge condition (\ref{residual}) 
does not
change the form of the equation (\ref{eq_v}). 
We can also verify this from the condition (\ref{residual})
straightforwardly as follows.
{}From Eq.(\ref{residual}) we find the identity 
${\rm e}^{-2A}\square_{D-1}\xi_2 =
{D-1 \over D-3}\partial_y\left(\partial_y+
(D-3)\dot{A}\right)\xi_2$. 
Combining this and
Eq.(\ref{delta_v}), we find 
$\delta v_\rho{^{,\rho}} = {2(D-2)(D-1) \over (D-3)^2} 
{\rm e}^{2A}\left(\partial_y+(D-1)\dot{A}\right) 
\partial_y
\left(\partial_y+(D-3)\dot{A}\right) \xi_2$ 
and $\partial_y\delta v_\mu =
{2(D-2) \over D-3}\partial_y
\left(\partial_y + (D-3)\dot{A}\right)\xi_{2,\mu}$. 
Hence, $\delta v_\mu$ satisfies just the same equation 
as Eq.(\ref{eq_v}) : 
\begin{eqnarray}
\left(\partial_y + (D-1)\dot{A}\right)
\partial_y\delta v_\mu 
= \frac{D-3}{D-1}{\rm e}^{-2A}
\delta v_\rho{^{,\rho}}{_{,\mu}}.
\end{eqnarray}
Therefore, $v_\mu$ can be
eliminated in the gauge, if we can set at a given $y=y_0$ surface
\begin{eqnarray}
v_\mu = - \delta v_\mu,\quad
\partial_y
v_\mu = - \partial_y
\delta v_\mu.\label{tt_condition}
\end{eqnarray} 
To clear matters, we introduce new functions
$\Lambda_\mu(x) \equiv \hat\xi_\mu(x,y_0)$, 
$\Xi_\mu(x) \equiv \left(\partial_y
\hat\xi_\mu\right)(x,y_0)$,
$\Gamma(x) \equiv \xi_2(x,y_0)$,
$\Delta(x) \equiv \left(\partial_y
\xi_2\right)(x,y_0)$ which
are defined at $y=y_0$ surface. In terms of these functions
Eq.(\ref{tt_condition}) can be rewritten as
\begin{eqnarray}
\square_{D-1}\Lambda_\mu 
+ \frac{D-3}{D-2}\Lambda_{\rho,\mu}{^{,\rho}} 
= \mathcal{A}_\mu,\quad
\square_{D-1}\Xi_\mu 
+ \frac{D-3}{D-2}\Xi_{\rho,\mu}{^{,\rho}} 
= \mathcal{B}_\mu,\label{const_surface}
\end{eqnarray}
where $\mathcal{A}_\mu(x) \equiv v_\mu(x,y_0)$ and
$\mathcal{B}_\mu(x) \equiv \left(\partial_y 
v_\mu\right)(x,y_0)$.
Similarly, the gauge condition (\ref{residual}) at $y=y_0$
surface can be rewritten as 
\begin{eqnarray}
\Xi_{\mu} + {\rm e}^{-2A}\Gamma_{,\mu} = 0,\quad
\Delta + (D-3)\dot{A}\Gamma 
= -\frac{D-3}{D-1}\Lambda_\rho{^{,\rho}}.\label{residual2}
\end{eqnarray}
$\Lambda_\mu$ and $\Xi_\mu$ can be determined similarly 
to the
Eq.(\ref{eq_v}). 
Next, we determine $\Gamma$ from the first equation of
Eq.(\ref{residual2}). 
However, this equation does not necessarily have a
solution for a general function $\Xi_\mu$. 
To see this in detail, plug this into the second
equation of Eq.(\ref{const_surface}) and we obtain 
$\square_{D-1}\Gamma_{,\mu} = -
\dfrac{D-2}{2D-5}{\rm e}^{2A}\mathcal{B}_\mu$. 
This equation can be solved if
and only if
$\mathcal{B}_\mu$ is expressed as a gradient of some 
function. 
In our case we obtain $\mathcal{B}_\mu = 2\partial_\mu D$
from Eq.(\ref{1st_gauge_3}). 
Hence, a solution $\Gamma$ of the first
equation of Eq.(\ref{residual2}) exists. 
At the end, $\Delta$
is determined from the second equation of 
Eq.(\ref{residual2}). 
In this
gauge, we obtain $D(x,y) = E{\rm e}^{-(D-1)A}$ where 
$E$ is a constant from
Eq.(\ref{1st_gauge_3}) and (\ref{1st_gauge_4}). 
We set
$E=0$ since we require that the fluctuations $B$ 
and $\varphi_{\rm R}$
should vanish at infinity $|x|\rightarrow\infty$ on the wall. 
Thus we established our gauge choice (Newton gauge) and the
constraints for the residual gauge transformations 
are (\ref{residual}) 
and 
\begin{eqnarray}
\square_{D-1}\hat\xi_\mu + \frac{D-3}{D-1}
\hat\xi_\rho{^{,\rho}}{_{,\mu}}=0.
\end{eqnarray}

%%%%%%%%%%%%%%%%%%%%%%%%%%%%%%%%%%%%%%%%%%%%%%%%%%%%%%%%%%%%%%%%%%%%%%%%%%%%%%%%%%%%%%%%%%%%%%%%%%

\section{Appendix B}

\begin{figure}[ht]
\begin{center}
\includegraphics[width=6.5cm]{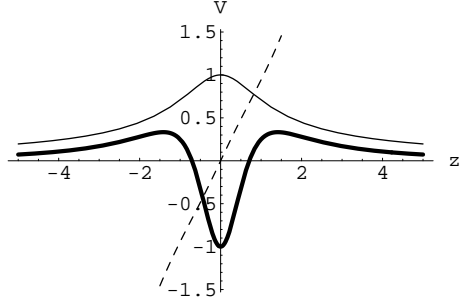}
 \caption{{\sf {\small The Schr\"odinger potential 
 for the tensor
 perturbation in terms of the coordinate system $z$. 
 We choose $k=\Lambda=1$. 
 The bold line denotes the potential, the solid line 
 denotes the graviton $\hat{h}^{\rm TT(+)}_{\mu\nu}$ 
 and the broken
 line denotes $\hat{h}^{\rm TT(-)}_{\mu\nu}$.}}}
\label{V_tensor_z}
\end{center}
\end{figure}

 For a special case where $k\Lambda^{-1}$ is an
integer\footnote{
Then we can no longer take the thin wall limit 
of $\Lambda\rightarrow \infty$ with $k$ fixed. 
}, we can 
express the Schr\"odinger potential in terms of 
$z$ explicitly. 
 As an illustrative example, 
 let us take 
$k=\Lambda$, where we find (putting $y_0=0$) 
${\rm e}^{A}=(\cosh ky)^{-1}$, $z=k^{-1}\sinh ky$, 
$A(z)=-\dfrac{1}{2}\log\left(1+k^2z^2\right)$. The Schr\"odinger
potential for the tensor perturbation takes the form : 
\begin{eqnarray}
\mathcal{V}_t(z) = - \frac{k^2(1-2k^2z^2)}{(1+k^2z^2)^2}.
\end{eqnarray}
There remains only one parameter controlling both 
the width of the 
wall and the magnitude of the gravitational coupling, 
similarly to Ref.\cite{Gremm:1999pj}.
Zero mode wave functions can also be expressed 
in terms of the $z$ coordinate explicitly and are shown in 
Fig.\ref{V_tensor_z} : 
\begin{eqnarray}
\tilde{h}^{\rm TT(0)}_{\mu\nu}(x,z) 
= \hat{h}^{\rm TT(+)}_{\mu\nu}(x)\ \frac{1}{\sqrt{1+k^2z^2}}
+ \hat{h}^{\rm TT(-)}_{\mu\nu}(x)\ 
\frac{z+k^2z^3/3}{\sqrt{1+k^2z^2}}.
\end{eqnarray}

%%%%%%%%%%%%%%%%%%%%%%%%%%%%%%%%%%%%%%%%%%%%%%%%%%%%%%%%%%%%%%%%%%
%%%%%%%%%%%%%%%%%%%%%%%%%%%%%%%%%%%%%%%%%%%%%%%%%%%%%%%%%%%%%%%%%%

\begin{figure}[t]
\begin{center}
\includegraphics[width=6.5cm]{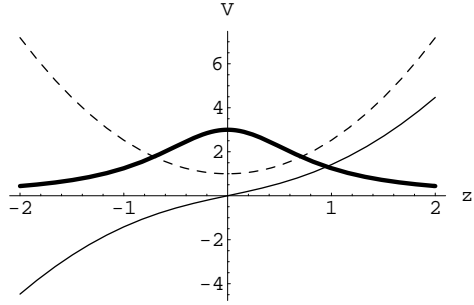}
 \caption{{\sf {\small The Schr\"odinger potential 
 for the scalar
 perturbation in terms of the coordinate system $z$. 
 We choose
 $k=\Lambda=1$. The bold line denotes the potential, 
 the solid line
 denotes $\hat{B}^{\rm (+)}$ and the broken 
 line denotes $\hat{B}^{\rm (-)}$.}}}
\label{V_scalar_z}
\end{center}
\end{figure}

The Schr\"odinger potential
$\mathcal{V}_e$ for the active scalar perturbation 
can be also expressed 
in terms of $z$ :  
\begin{eqnarray}
\mathcal{V}_e = 
\frac{k^2\left(3+2k^2z^2\right)}{\left(1+k^2z^2\right)^2},
\end{eqnarray}
and zero modes are of the form : 
\begin{eqnarray}
\hat{B}^{(0)}(x,z) = \hat{B}^{(+)}(x)\ z\sqrt{1+k^2z^2}
+ \hat{B}^{(-)}(x)\ \left(1+z\tan^{-1}kz\right)\sqrt{1+k^2z^2}.
\end{eqnarray}
These are shown in Fig.\ref{V_scalar_z}.
%%%%%%%%%%%%%%%%%%%%%%%%%%%%%%%%%%%%%%%%%%%%%%%%%%%%%%%%%%%%
%%%%%%%%%%%%%%%%%%%%%%%%%%%%%%%%%%%%%%%%%%%%%%%%%%%%%%%%%%%%

%%%%%%%%%
%%%%%%%%%%%%%%%%%%%% Reference 
%%%%%%

\end{document}